\documentclass[a4paper,UKenglish,cleveref, autoref, thm-restate]{lipics-v2021}

\pdfoutput=1 
\hideLIPIcs  

\title{Depth-Optimal Quantum Layout Synthesis as SAT\thanks{This is the technical report for
A. B. Jakobsen, A. B. Clausen, J. van de Pol and I. Shaik, {\em Depth-Optimal Quantum Layout Synthesis as SAT}.
 In: Proc. IC on Theory and Applications of Satisfiability Testing, {SAT} 2025,
(SAT'25), Glasgow, UK, 2025.}}

\titlerunning{Depth-Optimal Quantum Layout Synthesis as SAT}

\author{Anna B. Jakobsen}{Department of Computer Science, Aarhus University, Denmark}{ablumejakobsen@gmail.com}{0009-0005-7892-7230}{}
\author{Anders B. Clausen}{Department of Computer Science, Aarhus University, Denmark}{anbclausen@gmail.com}{0009-0008-3103-4365}{}
\author{Jaco van de Pol}{Department of Computer Science, Aarhus University, Denmark}{jaco@cs.au.dk}{0000-0003-4305-0625}{}
\author{Irfansha Shaik}{Department of Computer Science, Aarhus University, Denmark\\Kvantify Aps, DK-2300 Copenhagen S, Denmark}{irfansha.shaik@cs.au.dk}{0000-0002-7404-348X}{}

\authorrunning{Anna B. Jakobsen, Anders B. Clausen, Jaco van de Pol, and Irfansha Shaik}
  
\Copyright{Anna B. Jakobsen, Anders B. Clausen, Jaco van de Pol, and Irfansha Shaik}

  

\supplementdetails[subcategory={Github}, cite={}, swhid={}]{Software}{https://github.com/anbclausen/quills}
\supplementdetails[subcategory={Zenodo link}, cite={}, swhid={}]{Software and Experimental Data}{https://doi.org/10.5281/zenodo.15606051}

\funding{This research was partially funded by the Innovation Fund Denmark project
``Automated Planning for Quantum Circuit Optimization''.
This research was also partially funded by the European Innovation Council through Accelerator grant no. 190124924.}

\acknowledgements{This paper is based on the MSc-thesis of Anna B. Jakobsen and Anders B. Clausen. They contributed equally to this work. We thank Kostiantyn V. Milkevych for porting Quills to Q-Synth v4.}

\nolinenumbers 

\EventEditors{Jeremias Berg and Jakob Nordstr\"{o}m}
\EventNoEds{2}
\EventLongTitle{28th International Conference on Theory and Applications of Satisfiability Testing (SAT 2025)}
\EventShortTitle{SAT 2025}
\EventAcronym{SAT}
\EventYear{2025}
\EventDate{August 12--15, 2025}
\EventLocation{Glasgow, Scotland}
\EventLogo{}
\SeriesVolume{341}
\ArticleNo{18}

\usepackage{amsfonts}
\usepackage[noend]{algpseudocode}
\usepackage{algorithm}
\usepackage{graphicx}
\usepackage{listings}
\usepackage{textcomp}
\usepackage{hyperref}
\usepackage{xcolor, xspace}
\usepackage{quantikz}
\usepackage{listings}
\usepackage{rotating}
\usepackage{float}
\usepackage{enumitem}
\usepackage{subcaption}
\usepackage{todonotes}
\usetikzlibrary{automata, positioning, arrows}
\usepackage{newfloat}

\newcommand{\mc}[1]{\ensuremath{\mathcal{#1}}\xspace}

\newcommand{\Cc}{\mc{C}}

\newcommand{\Pc}{\mc{P}}

\newcommand{\I}[1]{\textit{#1}}

\newcommand{\U}[1]{\underline{#1}}
\newcommand{\T}[1]{\texttt{#1}}

\newcommand{\quills}{QuilLS}

\begin{document}

\maketitle

\begin{abstract}
Quantum circuits consist of gates applied to qubits.
Current quantum hardware platforms impose connectivity restrictions on binary CX gates. Hence, Layout Synthesis is an important step to transpile quantum circuits before they can be executed. Since CX gates are noisy, it is important to reduce the CX count or CX depth of the mapped circuits.

We provide a new and efficient encoding of Quantum-circuit Layout Synthesis in SAT. Previous SAT encodings focused on gate count and CX-gate count.
Our encoding instead guarantees that we find mapped circuits with minimal circuit depth or minimal CX-gate depth.
We use incremental SAT solving and parallel plans for an efficient encoding.
This results in speedups of more than 10-100x compared to OLSQ2, which guarantees depth-optimality. But minimizing depth still takes more time than minimizing gate count with Q-Synth.

We correlate the noise reduction achieved by simulating circuits after (CX)-count and (CX)-depth reduction.
We find that minimizing for CX-count correlates better with reducing noise than minimizing for CX-depth. However, taking into account both CX-count and CX-depth provides the best noise reduction.
\end{abstract}

\section{Introduction}
With recent advances in quantum hardware, there is a surge of interest in scalable compilation techniques.
In order to execute a quantum program, it must be compiled to a circuit with unary and binary gates, while handling hardware restrictions.
The quantum compilation pipeline broadly consists of two steps: Circuit Synthesis and Layout Synthesis.
In this paper, we focus on the Quantum Layout Synthesis (QLS) problem~\cite{sabre, Wille_2019, olsq, qsynth_sat} that transforms a ``logical'' quantum circuit consisting of native gates into an equivalent circuit that respects the connectivity of a physical quantum platform.
Many quantum platforms do not support all-to-all connectivity of physical qubits, thus restricting the application of binary gates to only neighboring qubits.
We assume that the only gates in the input circuit are unary gates and the binary CX (Conditional NOT) gate. The physical platform is specified by an (undirected) graph on physical qubits, the coupling map. The output is an initial mapping of the logical qubits onto the physical qubits, as well as a new, equivalent circuit, in which CX gates are only applied to physical qubits that are connected in the coupling map. To ensure this, so-called SWAP gates may be inserted to route interacting logical qubits towards connected physical qubits. During routing, the order of the gates must also be respected.

Note that the QLS problem may admit many possible solutions.
One can easily schedule all gates by eagerly adding SWAP gates to bring distant logical qubits closer together.
Although the primary objective of QLS is to handle connectivity restrictions, the secondary objective is to maximize hardware performance.
In the current Noisy Intermediate Scale Quantum (NISQ) era, qubits are noisy, and every gate adds to the error.
Reducing noise is crucial for practical quantum computing.
While there does not exist a single metric that predicts actual hardware performance, there exist some simple metrics like gate-count that are indicative.
Since binary CX gates can be up to 10 times more error-prone than unary gates, the number of CX gates (CX-count metric) is more relevant than just gate count~\cite{sqgm}.
Since SWAP gates are usually implemented by 3 CX gates each, solutions with too many SWAPs should be avoided.
Several heuristic approaches have been proposed that reduce the additional SWAP gates.
The most famous is SABRE~\cite{sabre} (available in IBM's Qiskit), which implements a very fast heuristic to find a correct layout mapping.
The tool MQT QMAP provides a heuristic approach based on the A*-algorithm~\cite{Zulehner_2019}.
There exist a myriad of other heuristic approaches that focus on CX-count optimization, we refer to~\cite{Tan2021} for a detailed survey.
While heuristic approaches are fast, their solutions are often far from optimal~\cite{olsq}.
Unfortunately, the optimal QLS problem is NP-hard~\cite{npc}.
The problem remains hard when considering natural restrictions on the coupling graph, like planar graphs~\cite{planar-hcp-is-npc} and bipartite graphs~\cite{bipartite-hcp-is-npc}, and it is even conjectured to be NP-hard for simple linear neighbor graphs~\cite{linear-neighbour-graph-npc-conjecture}. 
Thus, unless P=NP, one needs exhaustive exact approaches for (near)-optimal solutions.

Several exact approaches based on SMT~\cite{Wille_2019,olsq}, Classical Planning~\cite{qsynth_planning}, and SAT~\cite{qsynth_sat, yang_et_al:LIPIcs.SAT.2024.29} have been proposed to optimize CX-count.
Synthesizing a $k$-SWAP optimal circuit involves refuting the existence of a mapped circuit of at most $k-1$ SWAPs.
Recent SAT approaches proposed encodings whose makespan corresponds to the number of inserted SWAP gates instead of the number of CX-gates.
Since there are typically fewer SWAP gates than CX gates, such SAT encodings significantly outperform earlier approaches~\cite{qsynth_sat}.

Despite the progress in CX-count optimization, CX-count alone is not sufficient to predict the hardware performance.
Note that qubits become unstable over time due to so-called \emph{decoherence}.
It is essential to execute gates in the circuit before the qubits become unstable.
To reduce execution time, multiple gates can be executed in parallel, provided they are independent, i.e., do not act on the same qubits.
This makes circuit depth an important metric, since it captures this notion of parallel execution.
Since CX-gates are up to 10x slower compared to unary gates, we focus on reducing the CX-depth of the circuit. This metric counts the longest chain of dependent CX-gates.

Optimization of circuit depth has been explored in~\cite{olsq} in the context of SMT solving.
Their tool OLSQ2 is the state-of-the-art tool for depth-optimal synthesis.
In practice, optimization of (CX)-depth is harder than (CX)-count.
Unlike (CX)-count optimization, the makespan of an encoding optimizing (CX)-depth must be at least the (CX)-depth of the initial circuit.
Since the (CX)-depth is typically higher than the number of SWAPs, depth optimal synthesis becomes much harder in practice.
Compared to the SAT based state-of-the-art tool Q-Synth v2 for CX-count optimization, OLSQ2 is indeed up to 5 orders of magnitude slower~\cite{qsynth_sat,olsq}.
Even the best SMT-based near-optimal tool TB-OLSQ2 is up to 2 orders of magnitude slower compared to Q-Synth v2~\cite{qsynth_sat}.
SAT-based tools are shown to significantly outperform SMT-based tools for QLS~\cite{qsynth_sat, yang_et_al:LIPIcs.SAT.2024.29}.
The natural next step is to investigate if an efficient SAT-based depth optimal encoding is feasible.

\paragraph*{Contributions} In this paper, we provide an encoding of (CX)-depth-optimal QLS in SAT for the first time. Additionally, we provide a tool, \quills{}\footnote{Available at \url{https://github.com/anbclausen/quills}, also ported to Q-Synth v4 at \url{https://github.com/irfansha/Q-Synth}.}, which implements the encoding.
Our approach is based on increasing the time horizon, where in each time step one layer of independent gates is added.
The first SAT solution found corresponds to a mapped circuit with optimal depth.
We use incremental SAT solving (reusing partial information from previous calls)~\cite{CDCL2021} and parallel plans to encode depth as makespan for efficient solving.

A second contribution is an extensive experiment, comparing various tools with respect to their running time and their performance in actual (CX)-count and (CX)-depth metrics.
We compare our new tool with SABRE, Q-Synth and (TB)-OLSQ2.
We run these tools on a standard set of quantum circuit benchmarks, obtained from~\cite{olsq,qsynth_sat} and on the coupling graph of a number of standard quantum platforms (with 5-80 qubits).
Our tool is often more than 10-100x faster compared to previous tools with guaranteed depth-optimality, but still slower than size-optimal tools.

Finally, we investigate how the various optimization metrics correlate with a reduction in noise, which is the ultimate goal of optimal QLS. Following the approach in~\cite{id-gates, resynth-noise}, we simulate all mapped circuits on a quantum platform with and without simulated noise, using the Qiskit simulator, and compute the Hellinger distance~\cite{noise1,noise2,noise3} of these runs. Surprisingly, we find that a reduction in CX-count correlates better with noise reduction than CX-depth. However, taking both measures, CX-count + CX-depth, into account leads to the best noise reduction, motivating the importance of our contribution.

\begin{figure}[htbp]
\centering
    
\begin{subfigure}{0.5\textwidth}
\centering
\includegraphics[scale=0.45]{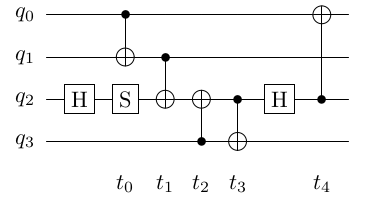}
\caption{An Example circuit.}
\label{fig:orgcircuit}
\end{subfigure}
\hfill
\begin{subfigure}{0.5\textwidth}
\includegraphics[scale=0.45]{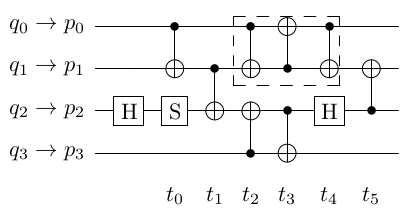}
\caption{CX-depth optimal circuit with 1 extra SWAP.}
\label{fig:mappedcircuit}
\end{subfigure}

\caption{Example circuit, before and after mapping on a linear neighbour graph. The dotted box represents the extra SWAP gate implemented as 3 CX gates. Time steps $t_0$ to $t_5$ indicate the CX gate parallel steps.}
\label{fig:example}
\end{figure}

\section{Preliminaries}
\subsection{Layout Synthesis of an Example Circuit}

As discussed previously in the introduction, the Quantum Layout Synthesis (QLS) problem takes as inputs a quantum circuit and a platform coupling map.
For example, consider the $4$-qubit circuit in Figure~\ref{fig:orgcircuit} with $5$ CX-gates and $3$ unary gates.
For simplicity, consider a quantum platform of 4 qubits $\{p_0,p_1,p_2,p_3\}$ with linear connections i.e., only $(p_0,p_1), (p_1,p_2), (p_2,p_3)$ are connected.
Notice that CX gates at $t_0, t_1, t_4$ form a triangle in Figure~\ref{fig:orgcircuit}, however our linear platform does not have a triangle.
Thus, there is no mapping from logical to physical qubits such that all the CX gates can be scheduled.
In such cases, SWAP gates can be inserted into the circuit, which allow physical qubits to swap their states. A SWAP gate can be constructed of three CX gates~\cite{Nielsen_2010}.
Indeed, as shown in Figure~\ref{fig:mappedcircuit}, by using one extra SWAP gate we can schedule all gates respecting connectivity restrictions.

Layout synthesis can be optimized using various metrics. One is size optimality, where the number of gates in the output circuit is minimized. In practice, this means minimizing the number of SWAP gates. Another is minimizing the depth of the output circuit, which is defined as the longest path in the dependency graph of the gates in the circuit. A similar metric is the CX-depth, where only the dependencies between the CX gates are considered.
Our original circuit has CX-depth $5$ where as the mapped circuit has optimal CX-depth of $6$.
In Figure~\ref{fig:example}, we indicate the parallel steps with CX-gates before and after mapping.
The number of parallel steps with CX-gates essentially corresponds to CX-depth.
Optimization of CX-count differs from CX-depth largely due to the interleaving of SWAP gates with other CX gates.
The choice of optimization goal can have a large impact. For example, one of the benchmark circuits used in our experimental evaluation (`\T{tof\_5}'), has a depth-optimal solution on the platform Guadalupe of depth 64 using 31 SWAP gates, while a size-optimal circuit uses only 5 SWAP gates, but has depth 71.

\subsection{Layout Synthesis as SAT}
\begin{figure}[htbp]
\centering
\includegraphics[scale=0.45]{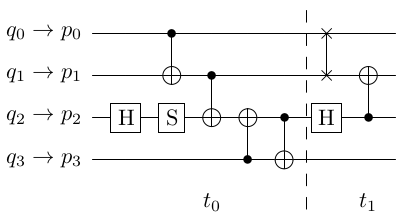}
\caption{Mapped circuit with only 2 parallel steps in the context of CX-count optimization.}
\label{fig:cnotoptexample}
\end{figure}

Given a propositional formula, the SAT problem is to find an assignment to the variables such that the formula evaluates to true.
While the general problem is NP-complete, in practice several instances can be solved efficiently by modern SAT solvers~\cite{HandbookSAT}. This motivates encoding other NP-complete problems to SAT. However, finding an efficient problem encoding is highly non-trivial.
As discussed earlier, recent SAT based approaches~\cite{qsynth_sat, yang_et_al:LIPIcs.SAT.2024.29} perform well on CX-count metric.
Unlike (CX)-depth optimization, a SAT encoding optimizing (CX)-count only requires a makespan of 2 on the example in Figure~\ref{fig:cnotoptexample}.
The key idea is to group CNOTs at each parallel step and add a single SWAP gate from time step $t_1$.
Thus, the makespan scales with the number of SWAPs which is typically much lower than the (CX)-count or (CX)-depth.

While the disadvantage in the larger required makespan is unavoidable, we can still apply parallel plans~\cite{DBLP:conf/kr/KautzMS96} to schedule parallel gates in a mapped circuit.
It is possible to encode depth in a sequential encoding, however allowing independent gates in the same time step can reduce the makespan. The makespan refers to the circuit (CX)-depth of the solution.
Reducing makespan indeed avoids many variable copies, thus impacting performance.
We adapt another main technique, incremental SAT solving, used in the earlier SAT encodings for our (CX)-depth optimal encoding.
Remember that an optimal synthesis of $k$ (CX)-depth circuit requires refuting the existence of $k-1$ or less (CX)-depth circuits.
Instead of checking for circuits with (CX)-depth from $0$ to $k-1$ independently, we reuse information from earlier SAT calls.
By using incremental SAT solving, we can maintain a single CNF formula and ``turn clauses on or off'' instead of rebuilding the entire CNF formula every time. This also increases the performance during \I{inprocessing} (the actual work of the SAT solver). 
Finally, for a (CX)-depth optimal encoding, we need to handle the execution of SWAP gates over 3 time steps instead of just 1.

\section{Depth-optimal SAT encoding}
\label{sec:sat-encoding}

In the previous sections, we observed that SAT-based encodings in Q-Synth perform very well in the case of size-optimal synthesis.
In this section, we formulate a similar SAT encoding of depth-optimal layout synthesis. In contrast to size-optimal synthesis, depth-optimal synthesis might produce solutions with a sub-optimal number of CX gates to find a solution with optimal circuit depth. Circuit depth refers to the longest path in the gate dependency graph. 
We employ two techniques, parallel plans, and incremental SAT solving for an efficient encoding.
Unlike Q-Synth, a parallel step in our encoding encodes a single layer in a circuit, consisting of parallel gates applied to different qubits.
Thus, the makespan of the encoding corresponds to the depth of the mapped circuit.
The novelty of our encoding comes from the handling of SWAP gates while maintaining depth-optimality, which is non-trivial.

\renewcommand{\algorithmicrequire}{\textbf{Input:}}
\renewcommand{\algorithmicensure}{\textbf{Output:}}
    
\begin{algorithm}[b]
\caption{Building and solving the SAT encoding.}
    \label{fig:sat-pseudo}
    \begin{algorithmic}
    \Require{Coupling map $\Pc$ and circuit $\Cc$}
    \State let $D$ be the depth of $\Cc$
    \State let instance be an empty CNF
    \For{$t = 1, 2, \dots$}                   
        \State Add $MappingConstraints(t)$ to instance
        \State Add $ConnectivityConstraints(t)$ to instance
        \State Add $GateConstraints(t)$ to instance
        \State Add $SwapConstraints(t)$ to instance
        \If{$t \geq D$}
            \State Add $Assumptions(t)$ to instance
            \State Solve instance with assumption $asm^t$
            \If{instance is satisfied}
                \State \Return {satisfying assignment} 
            \EndIf
        \EndIf
    \EndFor
    \end{algorithmic}
\end{algorithm}

The overall structure of the encoding can be seen in Algorithm \ref{fig:sat-pseudo}. 
Since the output cannot have smaller depth than the input, we only test the existence of a mapping when the makespan is at least as large as the input depth. The first time step where the instance is satisfied is the optimal depth.
Much of the structure of the encoding follows that of~\cite{qsynth_sat}. We will focus the in-depth explanation to those parts that are different.

\begin{table}[tbp]
  \centering
  \caption{Encoding variables and descriptions}
  \begin{tabular}{ll}
  \hline
  Variable & Description \\
  \hline
    $\text{mp}^t_{q,p}$ & logical qubit $q$ is mapped to the physical qubit $p$ at time $t$.\\
    $\text{oc}^t_p$ & physical qubit $p$ is occupied by some logical qubit at time $t$.\\
    $\text{e}_{q,q'}^{t}$ & logical qubits $q, q'$ are mapped to connected qubits at time $t$.\\
    $\text{u}_{p}^{t}$ & physical qubit $p$ is not part of a SWAP gate at time $t$.\\
    $\text{c}^t_g$ & gate $g$ is currently applied at time $t$.\\
    $\text{a}^t_g$ & gate $g$ was applied at a time strictly before $t$.\\
    $\text{d}^t_g$ & gate $g$ is delayed and has not been applied yet at time $t$.\\
    $\text{sw}^t_{p,p'}$ & physical qubits $p, p'$ are swapped at times $t$, $t-1$ and $t-2$.\\
    $\mathit{Sw(p)}^t$ & the set of variables $\text{sw}^t_{p,p'}$ where $(p,p') \in \mathit{P_{conn}}$\\
    $\text{st}^t_{p}$ & physical qubit $p$ is touched by a SWAP at $t$, $t-1$ and $t-2$.\\
    $\text{asm}^t$ & assumption to indicate that $t$ is the final time step.\\
  \hline
  \end{tabular}
  \label{table:vars}
  \end{table}

Table~\ref{table:vars} presents the variables used in the encoding.
For each time step, a set of variables is generated for the constraints of that time step. The asm$^t$ variable is special and is used for incremental SAT solving; it provides context to the SAT solver about previous attempts to find a solution, essentially reducing the search space of the solver. SWAP gates are scheduled over 3 time steps since they are commonly implemented as three consecutive CX gates. In the following, we describe each part of the constraints added for each time step.

\noindent
The Mapping and Connectivity constraints are exactly as in~\cite{qsynth_sat}, but are repeated here for the sake of completeness. 
We use $P$ for the set of physical qubits and $Q$ for the set of logical qubits. $P_{\mathit{conn}}$ indicates the edges between physical qubits in the connectivity graph.
We use $g.q$ to indicate the logical qubit that a unary gate must be applied on, and $g.q_0, g.q_1$ for the qubits that a CX gate must be applied on. $\mathit{G_{CX}}$ denotes the set of CX gates in the input circuit while $\mathit{G_{unary}}$ denotes the set of unary gates.

\hfill

\noindent
\I{\U{MappingConstraints}}
\begin{gather}
    \bigwedge_{q \in Q} \text{ExactlyOne}(\text{mp}^t_{q,p_0}, \dots, \text{mp}^t_{q,p_n}) \label{eq:exactly-one}\\
    \bigwedge_{p \in P} \text{AtMostOne}(\text{mp}^t_{q_0,p}, \dots, \text{mp}^t_{q_m,p}) \label{eq:atmost-one}\\
    \bigwedge_{p \in P} \Big( \text{oc}^t_p \iff \big( \bigvee_{q \in Q} \text{mp}^t_{q,p}\big) \Big) \label{eq:occupied}
\end{gather}

The above constraints ensure that the mp$^t_{q, p}$ and oc$^t_p$ variables are set correctly. Here ExactlyOne$(t_1, \ldots, t_n) \triangleq \sum_{i=1}^n t_i = 1$ denotes that exactly one of $\{t_1, \ldots, t_n\}$ is true while AtMostOne$(t_1, \ldots, t_n) \triangleq \sum_{i=1}^n t_i \leq 1$ denotes that at most one of $\{t_1, \ldots, t_n\}$ is true.

\hfill

\newpage
\I{\U{ConnectivityConstraints}}

\begin{gather}
    \bigwedge_{q,q' \in Q} \Big( 
    \bigwedge_{(p, p') \in P_{\mathit{conn}}} \big( (\text{mp}^t_{q,p} \land \text{mp}^t_{q',p'}) 
        \Rightarrow \text{e}^t_{q,q'} \big)\quad \land 
    \bigwedge_{p, p' \not\in P_{\mathit{conn}}} \big( (\text{mp}^t_{q,p} \land \text{mp}^t_{q',p'}) 
        \Rightarrow \neg\text{e}^t_{q,q'} \big) \Big) \label{eq:enabled}\\
    \bigwedge_{g \in \mathit{G_{CX}}} \big( \text{c}^t_g \implies \text{e}^t_{g.q_0, g.q_1} \big) \label{eq:current-is-enabled}
\end{gather}

Constraint \ref{eq:enabled} ensures that the $e^t_{q, q'}$ predicates are true if and only if the physical qubits assigned to $q$ and $q'$ at time $t$ are connected. Constraint \ref{eq:current-is-enabled} ensures that the CX gates are applied to connected physical qubits.

\hfill

\noindent
\I{\U{GateConstraints}}

\noindent
The Gate constraints follow a similar pattern to those in~\cite{qsynth_sat}, but are substantially different because in this encoding at most one gate can be scheduled on each qubit at a given time step.
For a given time $t$, a gate is either currently scheduled denoted $\text{c}^t_g$, it has been scheduled for some $t'<t$ denoted $\text{a}^t_g$ (advanced), or it will be scheduled for some $t' > t$ denoted $\text{d}^t_g$ (delayed) as encoded in constraint \ref{eq:gate-in-some-state}. Furthermore, no gate can be `advanced' on the first time step since nothing has been applied yet as encoded in constraint \ref{eq:gate-t1-not-advanced}.

\begin{gather}
    \bigwedge_{g \in G} \text{ExactlyOne}(\text{c}^t_g, \text{a}^t_g, \text{d}^t_g) \label{eq:gate-in-some-state}\\
    \bigwedge_{g \in G} \neg\text{a}^1_g \label{eq:gate-t1-not-advanced}
\end{gather}

The main idea from the size-optimal encoding was that of two-way propagation. This is adapted to a depth-optimal encoding by ensuring that if a gate is `current' or `delayed' its direct successors must be `delayed', while if a gate is `current' or `advanced' its direct predecessors must be `advanced'.
This information is propagated throughout the circuit, giving rise to the name.
Constraint \ref{eq:gate-dependencies} enforces that the dependencies of the input circuit are preserved in the output circuit and that two gates are not applied on the same qubit at the same time.
We use $S(g)$ (and $P(g)$) to denote the successors (predecessors) of gate $g$ in the dependency graph of the input circuit.
\begin{gather}
\bigwedge_{g \in G} \big( \bigwedge_{g' \in S(g)} (\text{c}^t_g \lor \text{d}^t_g) \implies \text{d}^t_{g'} \ \land \bigwedge_{g' \in P(g)} (\text{c}^t_g \lor \text{a}^t_g) \implies \text{a}^t_{g'}\big) \label{eq:gate-dependencies}
\end{gather}

To ensure that gates are not skipped or applied twice, constraint \ref{eq:not-skipped-or-applied-twice} ensures that a gate is `advanced' if and only if it was `current' or `advanced' in the previous time step. Likewise, a gate is `current' or `delayed' if and only if it was `delayed' in the previous time step.
\begin{gather}
    \bigwedge_{g \in G} \big(\big( (\text{c}^{t-1}_{g} \lor \text{a}^{t-1}_{g}) \iff \text{a}^{t}_{g} \big) \land \big( \text{d}^{t-1}_{g} \iff (\text{c}^{t}_{g} \lor \text{d}^{t}_{g}) \big)\big) \label{eq:not-skipped-or-applied-twice}
\end{gather}

Constraint \ref{eq:propagating-current} is added to speed up solving time by reducing the search space. This constraint express that when a gate is `current' none of its predecessors and successors can be `current'. We use $\mathit{fullS(g)}$ (and $\mathit{fullP(g)}$) to denote the transitive closure of the successor (predecessor) relation in the dependency graph. We do this to improve the automatic propagation of the dependency constraints.
\begin{gather}
    \bigwedge_{g \in G} \bigwedge_{g' \in \mathit{fullS(g)} \cup \mathit{fullP(g)}} \text{c}^t_g \implies \neg\text{c}^t_{g'} \label{eq:propagating-current}
\end{gather}

Whenever a gate is `current', the physical qubit(s) that it is being applied on must be `usable', i.e., not undergoing a SWAP. This is encoded for binary gates in constraint \ref{eq:usable} and for unary gates in constraint \ref{eq:usable-unary}. These constraints ensure that a gate and a SWAP are not applied at the same time step. 
\begin{gather}
    \bigwedge_{g \in \mathit{G_{CX}}} \bigwedge_{p \in P} \text{c}^t_g \implies ( (\text{mp}^t_{g.q_0, p} \lor \text{mp}^t_{g.q_1, p}) \implies \text{u}^t_p ) \label{eq:usable}\\
    \bigwedge_{g \in \mathit{G_{unary}}} \bigwedge_{p \in P} \text{c}^t_g \implies ( \text{mp}^t_{g.q, p} \implies \text{u}^t_p) \label{eq:usable-unary}
\end{gather}

\hfill

\noindent
\I{\U{SwapConstraints}}

\noindent
Constraint \ref{eq:physical-can-only-be-in-one-swap} encodes that a given physical qubit can participate in at most one SWAP gate for any three consecutive time steps. This is to ensure that SWAP gates are not applied in conflict with each other. We use $\mathit{Sw(p)}^t$ to mean the set of variables $\text{sw}^t_{p,p'}$ where $(p,p') \in \mathit{P_{conn}}$.
\begin{gather}
    \bigwedge_{p\in P} \text{AtMostOne}(\mathit{Sw(p)}^t \cup \mathit{Sw(p)}^{t-1} \cup \mathit{Sw(p)}^{t-2}) \label{eq:physical-can-only-be-in-one-swap}
\end{gather}

The `swap' variables indicate the last time step of a SWAP gate. To ensure that gates are not scheduled on top of SWAPs we require that if the `swap touched' variable is set for a qubit, it is not available for gates to be scheduled on it for this and the preceding two time steps as seen in constraint \ref{eq:swapping-qubits-are-busy}. Also, as encoded in constraint \ref{eq:cannot-swap-in-beginning}, the `swap' variables must all be set to false for the first three time steps.
\begin{gather}
    \bigwedge_{p \in P} \bigwedge_{t' \in \{t, t-1, t-2\}} \big( \text{st}^t_{p} \implies \neg\text{u}^{t'}_{p} \big) \label{eq:swapping-qubits-are-busy}\\
    \bigwedge_{p,p' \in P_{\mathit{conn}}} \left( \neg\text{sw}^1_{p, p'} \land \neg\text{sw}^2_{p, p'} \land \neg\text{sw}^3_{p, p'} \right) \label{eq:cannot-swap-in-beginning}
\end{gather}

Nothing stops the SAT solver from swapping unmapped physical qubits. Such SWAPs are referred to as 'spurious SWAPs'. To ensure that no spurious SWAPs are scheduled, we require that at least one physical qubit in a SWAP must have a logical qubit mapped to it in constraint \ref{eq:no-spurious-swaps}. We denote a SWAP with only one mapped physical qubit 'ancillary SWAP'. Ancillary SWAPs can be disabled by requiring that both physical qubits participating in a SWAP be occupied. We only present results allowing ancillary SWAPs.
\begin{gather}
    \bigwedge_{p, p' \in P_{\mathit{conn}}} \left( \text{sw}^t_{p,p'} \implies (\text{oc}^t_p \lor \text{oc}^t_{p'}) \right) \label{eq:no-spurious-swaps}
\end{gather}

Three final constraints are added to encode the effect of a SWAP gate. Constraint \ref{eq:swap-touch} says that if a `swap' variable is set which includes the physical qubit $p$, the `swap touched' variable for $p$ is also set. Constraint \ref{eq:mapping-preserved-if-no-swap} says that if the swap touched variable is not set for a qubit, its mapping is preserved from the last time step. Finally, constraint \ref{eq:swaps} says that when the swap variable for two qubits is set, the mappings for them switch compared to the previous time step.
\begin{gather}
    \bigwedge_{p \in P} \big( \text{st}^t_p \iff \big( \bigvee_{p' \in \mathit{conn(p)}} \text{sw}^t_{p,p'} \big) \big) \label{eq:swap-touch}\\
    \bigwedge_{p \in P} \bigwedge_{q \in Q} \left( \neg\text{st}^t_p \implies (\text{mp}^{t-1}_{q,p} \iff \text{mp}^{t}_{q,p}) \right) \label{eq:mapping-preserved-if-no-swap}\\
    \bigwedge_{p, p' \in P_{\mathit{conn}}} \bigwedge_{q \in Q} \Big( \text{sw}^t_{p, p'} \implies \big( 
        (\text{mp}^{t-1}_{q,p} \iff \text{mp}^{t}_{q,p'}) \land
        (\text{mp}^{t-1}_{q,p'} \iff \text{mp}^{t}_{q,p}) \big) \Big) \label{eq:swaps}
\end{gather}

\hfill

\noindent
\I{\U{Assumptions}}

\noindent 
We also encode an assumption for incremental SAT solving in constraint \ref{eq:assumption}: For each $t$ exceeding the depth of the input circuit, we add this constraint expressing that no gate is delayed at the last time step, 
which ensures that all gates from the input circuit are scheduled. This tells the SAT solver that we expect a solution at time step $t$, and consequently not at any time step prior to $t$.
\begin{gather}
    \bigwedge_{g \in G} \left( \text{asm}^t \implies \neg\text{d}^{t}_g \right) \label{eq:assumption}
\end{gather}

\subsection{Encoding variations}
\label{sec:enc-variations}

There are two goal variations that are easily implemented. The first one is finding optimal CX-depth instead of depth. This can be achieved by the same encoding, by preprocessing the circuit to only include CX gates. After the optimal placement of SWAPs has been found using the encoding, the unary gates can be reinserted to get the final circuit.

The other variation is finding a circuit with minimal depth and locally minimal SWAP count. After a circuit with minimal depth has been found, we count the number of SWAPs $S$ and force it to decrease in increments of one by adding new assumption variables $\text{sasm}^i$ for $i = S-1, \dots, 0$.
To force the number of SWAPs to be at most $i$, the following constraint is added to the instance: $\text{sasm}^i \implies \text{AtMostN}(i, \text{sw}^0_{p_0, p_1}, \dots, \text{sw}^t_{p_{n-1}, p_n})$, where $\text{sw}^0_{p_0, p_1}, \dots, \text{sw}^t_{p_{n-1}, p_n}$ represents all `swap' variables that have been created in total over all layers.
The first $i$ where it is not possible to solve an instance, tells us that the locally minimal number of SWAP gates is $i+1$. This is because the search space is monotone in the number of SWAP gates~\cite{olsq}.

\subsection{Count Optimal Encoding vs Depth Optimal Encoding}
\label{subsec:countvsdepthencoding}

We have employed several effective techniques from SAT based SWAP-count optimal synthesis in~\cite{qsynth_sat} for a scalable depth optimal encoding in this paper.
In this Section, we summarize the similarities and differences between the count optimal and depth optimal encodings.

The following list describes the techniques adapted from count optimal encoding:
\begin{itemize}
    \item Incremental solving to reuse learned clauses from earlier bounds with simple assumptions.
    \item Bi-directional propagation of gate constraints using `delayed' and `advanced' variables.
    \item Similar cardinality and connectivity constraints in both encodings.
\end{itemize}

The following list summarizes the main differences from the count optimal encoding:
\begin{itemize}
    \item A SWAP gate (with depth 3) can interleave with other 1-qubit and 2-qubit gates in depth optimal encoding unlike a count optimal setting. Equations $13$ to $19$ present  the new constraints for enabling interleaving of SWAP gates.
    \item The makespan of a count optimal encoding corresponds to the number of SWAPs needed. However, the makespan of a depth optimal encoding is at least the depth of the original circuit. Since the circuit depth/CNOT-depth is often several times more than the number of SWAPs, scalability of encodings differs greatly.
    \item As discussed in Section~\ref{sec:enc-variations}, a depth optimal encoding also enables local SWAP count optimization with simple additional constraints. On the other hand, one cannot easily optimize local depth in a count optimal encoding.
\end{itemize}

\section{Experimental evaluation}
Our SAT encoding is implemented in the tool \quills{}\footnote{\quills{} is available at \url{https://github.com/anbclausen/quills}. It is also ported to Q-Synth v4 at \url{https://github.com/irfansha/Q-Synth}. The experiments reported here are obtained with \quills{}.}.
Remember that optimizing depth is harder than optimizing count due to larger makespan.
For an evaluation, we are interested in the performance penalty by depth optimization instead of count.
Further, we are also interested in the quality of results and their impact on noise reduction which is the ultimate goal for circuit optimization.
In particular, we structure our experiments to answer the following research questions:
\begin{enumerate}[label=\Alph*]
    \item How does the runtime performance of \quills{} compare to Q-Synth, TB-OLSQ2, and OLSQ2?
    \item How do the depth, CX-depth, SWAP count of \quills{} compare to SABRE, Q-Synth, TB-OLSQ2, and OLSQ2?
    \item How do different optimization strategies affect the quality of the synthesized circuits in terms of noise?
\end{enumerate}
Our experiments are run on a Linux machine with a 2.4GHz CPU and 60GB of memory. We always use a timeout of 2 hours (7200 seconds). For our benchmarks, we use a set of 33 standard benchmarks (15 arithmetic and 18 QUEKO \cite{queko}) with up to 54 qubits (q) and 270 CX gates from the literature~\cite{olsq, qsynth_sat}. Furthermore, we use 10 8q Variational Quantum Eigensolver (VQE) circuits with up to 79 CX gates from~\cite{qsynth_sat} representing practically applicable quantum circuits.
Each of these benchmark circuits is synthesized on the following platforms: Tenerife (5q), Melbourne (14q), Guadalupe (16q), Tokyo (20q), and Cambridge (32q) by IBM; Google's Sycamore (54q); and Rigetti's Aspen-M (80q).
Given a test circuit, we only synthesize it on a platform if the platform has the required number of qubits. The exact data on the circuits can be found in \cref{app:Benchmarks}. 

For our evaluation, we use the following tools:
\begin{itemize}
    \item \quills{}: Our (CX)-depth-optimal tool, with CaDiCaL (v1.5.3)~\cite{BiereFazekasFleuryHeisinger-SAT-Competition-2020-solvers} SAT solver as our backend.
    \item SABRE~\cite{sabre}: A heuristic algorithm that is part of Qiskit. We use the same experimental setting as in~\cite{qsynth_sat}.
    \item Q-Synth (v2)~\cite{qsynth_sat}: A SAT based state-of-the-art size-optimal tool with CaDiCal (v1.5.3) as backend.
    \item OLSQ2~\cite{olsq}: An SMT-based state-of-the-art depth-optimal tool with Z3 (v4.13.0.0)~\cite{DBLP:conf/tacas/MouraB08} as a backend. 
    \item TB-OLSQ2~\cite{olsq}: A near-optimal version of OLSQ2 with better runtime performance. We report a timeout if the tool does not terminate within the time limit.
\end{itemize}
For both OLSQ2 and TB-OLSQ2, we use the best settings, i.e., allowing them to use SABRE to get an upper bound.

All measurements are available on Zenodo~\cite{jakobsen_2025_15606051}\footnote{Full code and data is available at \url{https://doi.org/10.5281/zenodo.15606051}} and an overview of the data is given in \cref{app:runtimedata}. Next, we will discuss the setup and the results of each experiment in detail.

\subsection{Performance of \quills{} Compared to Existing Tools}
\label{sec:runtime-performance}

\paragraph*{Experimental setup}
We are interested in both depth and CX-depth optimization.
As mentioned in Section \ref{sec:enc-variations}, we can optimize either depth or CX-depth in our SAT encoding.
We also ensure that \quills{} always computes locally minimal SWAPs.
The OLSQ2 and TB-OLSQ2 tools optimize depth, and we can make them optimize CX-depth instead, by removing all unary gates from the input circuit before passing it to these tools. For a fair comparison, we set options such that OLSQ2 and TB-OLSQ2 also compute locally minimal SWAP count.
We run SABRE, Q-Synth and the (CX)-depth-optimal versions of (TB)-OLSQ2 and \quills{} on all the benchmarks on every platform and measure running time. 

\begin{figure}[h]
    \centering
    \includegraphics[trim={3px 3px 3px 3px},clip,width=0.8\linewidth]{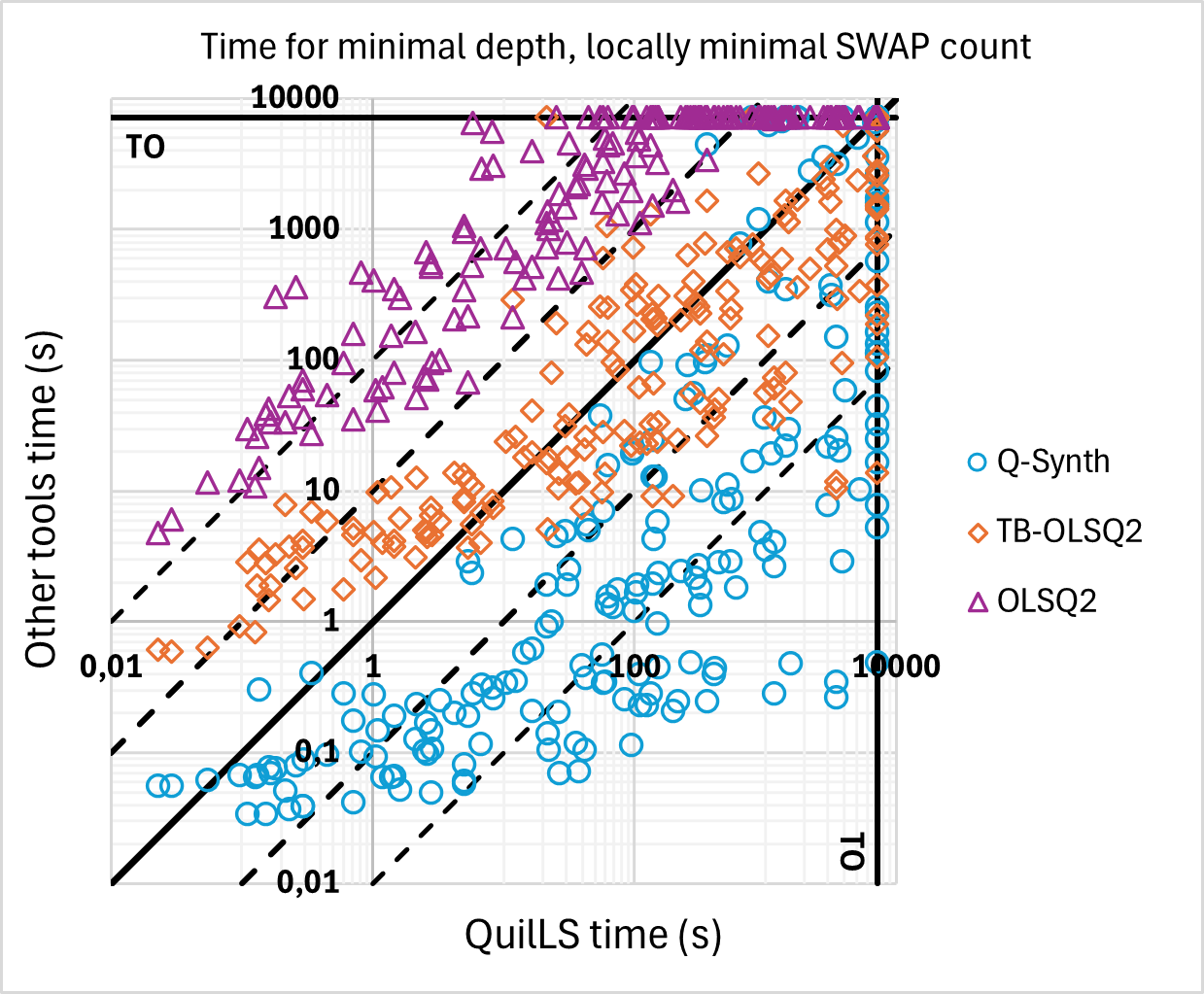}
    \caption{Time comparison between \quills{} and other tools (with their own optimization metrics).
    A mark on the bold diagonal line shows equal performance between the tools, a mark above shows better performance by \quills{}, and a mark below shows worse performance by \quills{} than the other tool.
    Dotted diagonal lines indicate performance ratios of 10 and 100.
    Bold horizontal and vertical lines indicate timeouts (TO).
    }
    \label{fig:min-depth-swap-times}
\end{figure}

 \paragraph*{Results}
First, we focus on depth optimization for tools \quills{} and OLSQ2.
The times for computing a heuristic solution with SABRE are excluded since they are all smaller than 0.5 seconds.
Figure \ref{fig:min-depth-swap-times} shows the running times comparison of \quills{} and the other tools. The full data set with running times for all tools can be found in the appendix.

In Table \ref{fig:tool-timeouts} the number of solved instances, and time-outs are shown for each tool/metric.
We also include the results with CX-depth optimization by \quills{} and OLSQ2.

\begin{table}[h]
\caption{Timeouts (TO), and solved instances for QLS tools.
}
\label{fig:tool-timeouts}
\centering
\begin{tabular}{|l|l|l|l|l|}
\hline
\textbf{Tool}       & \textbf{TO} & \textbf{Solved} & \textbf{Solved \%} \\ \hline
SABRE               & 0           & 220             & 100\%              \\ \hline
TB-OLSQ2 (CX-depth) & 26          & 194             & 88\%               \\ \hline
TB-OLSQ2 (depth)    & 30          & 190             & 86\%               \\ \hline
Q-Synth (SWAP count)            & 31            & 189             & 86\%               \\ \hline
QuilLS (CX-depth)   & 35          & 185             & 84\%               \\ \hline
QuilLS (depth)      & 48          & 172             & 78\%               \\ \hline
OLSQ2 (CX-depth)    & 102         & 118             & 54\%               \\ \hline
OLSQ2 (depth)       & 125         & 95              & 43\%               \\ \hline
\end{tabular}
\end{table}

\paragraph*{\quills{} vs OLSQ2 and TB-OLSQ2}
OLSQ2 is our direct competitor that optimizes the same metric, i.e., depth and locally minimal SWAP count.
From the results, it is clear that we significantly outperform OLSQ2.
OLSQ2 is almost always at least 10 times slower than \quills{}, and sometimes up to 3 orders of magnitude slower.
TB-OLSQ2 on the other hand, being a near-optimal tool performs better on circuits with larger synthesis times.
This is consistent with the observations between optimal and near-optimal synthesis in the literature~\cite{olsq}.
\paragraph*{\quills{} vs Q-Synth}
Q-Synth is generally 10 times faster than \quills{} and more than 100 times faster on many circuits.
Despite being similar encodings and using the same SAT solver as a backend, we see the hardness of optimizing depth compared to SWAP count.
Interestingly, on some circuits with large synthesis times Q-Synth uses more time than \quills{}.
\paragraph*{Depth vs CX-depth optimization in \quills{}}
It is faster in general to compute a CX-depth-optimal circuit than a depth-optimal circuit. This is expected, since the instances are smaller. This is also the case for (TB)-OLSQ2, and the performance for CX-depth-optimal synthesis gives a picture very similar to Figure \ref{fig:min-depth-swap-times}.

Q-Synth, TB-OLSQ2, and the CX-depth-optimal version of \quills{} solve 84-88\% of instances, while the depth-optimal version of \quills{} solves 78\% of instances. OLSQ2 solves around 50\% of instances.
SABRE solves every instance within a second.
However, in the next experiment we will see that the quality of mappings from SABRE are indeed far from optimal.

\subsection{Quality of Results for \quills{} Compared to Existing Tools}
\label{sec:result-comparison}

In the previous experiment, we have seen a performance penalty on optimizing depth instead of SWAP count.
In this experiment, we look at the quality of the results reported by all the tools.
We are interested in finding a trade-off between performance and quality of results.
So far, we have considered the two metrics SWAP count and depth with locally optimal SWAP count.
Since it is quite cheap to compute SWAP optimal mappings, one could also consider optimizing depth for the optimal SWAP count.
To do this, one can use e.g.\ Q-Synth to find the optimal SWAP count, and then use our tool \quills{} for optimizing depth given a bound on SWAP count (we call this combination `Q-S+Qui').
The experimental setup is the same as for the previous section, except that we include this new variant: size optimal with locally minimal (CX)-depth. We now measure the depth, CX-depth, and SWAP count for each synthesized circuit.

\paragraph*{Results}
Table \ref{fig:tool-ratios} shows for each tool and optimization metric the average ratio of the depth, CX-depth and CX count compared to optimal tools, over a set of benchmarks that all tools could solve (93 for all tools, 129 if we discard OLSQ2). We also show the success rate (number of solved circuits over all 220 benchmarks).
We report the number of CX gates in the synthesized circuit rather than the SWAP gates, since the CX gate count also reflects the size of the original circuit. 
We set the  tool and metric combination that guarantees optimal values to \textbf{1.00} and provide the ratios of the results from other tools.
A higher ratio means a worse result.

\begin{table}[h]
    \caption{Average ratios between depth (d), CX-depth (CX-d) and CX gate count (CX) of circuits mapped by various tool/metric combinations. Bold numbers indicate guaranteed optimality.
    93 circuits are solved by all tools; 129 circuits are solved by all tools except OLSQ2. 
    }
\label{fig:tool-ratios}
\centering
    \begin{tabular}{l@{ }l@{}r|lll|lll}
    \hline
    & & & \multicolumn{3}{c|}{Avg. over N=129} & \multicolumn{3}{c}{Avg. over N=93}\\
    \hline
    \multicolumn{3}{l|}{Tool/metric/\%success} & d. & CX-d & CX & d. & CX-d & CX\\
    \hline
    SABRE   & heur & 100 & 1.48  & 1.46       & 1.15       & 1.29 & 1.27 & 1.10 \\ \hline
    TB-O2   & CX-d &  88 & 1.09  & 1.06       & 1.01       & 1.08 & 1.06 & 1.01 \\ \hline
    TB-O2   & d.   &  86 & 1.08  & 1.07       & 1.01       & 1.08 & 1.07 & 1.01 \\ \hline
    Q-Synth & CX   &  86 & 1.06  & 1.05       & \textbf{1.00} & 1.04 & 1.03 & \textbf{1.00} \\ \hline
    QuilLS  & CX-d &  84 & 1.03  & \textbf{1.00} & 1.04    & 1.02 & \textbf{1.00} & 1.03 \\ \hline
    QuilLS  & d.   &  78 & \textbf{1.00} & 1.03  & 1.09    & \textbf{1.00} & 1.02 & 1.04 \\ \hline
    Q-S+Qui & CX-d &  67 & 1.04  & 1.01       & \textbf{1.00} & 1.03 & 1.01 & \textbf{1.00}\\ \hline
    Q-S+Qui & d.   &  60 & 1.02  & 1.02       & \textbf{1.00} & 1.01 & 1.02 & \textbf{1.00} \\ \hline
    OLSQ2   & CX-d &  54 & --  & --           & --         & 1.02 & \textbf{1.00} & 1.03 \\ \hline
    OLSQ2   & d.   &  43 & --  & --           & --         & \textbf{1.00} & 1.01 & 1.04 \\ \hline
    \end{tabular}
\end{table}

\paragraph*{Discussion}
For all three measures, SABRE gets the worst ratios, which is to be expected since SABRE is a heuristic tool; it is also the only tool that can map all instances. Comparing N=129 vs N=93, it is also clear that for harder instances, SABRE inserts relatively more SWAP gates.

Only OLSQ2 and QuilLS guarantee minimal (CX)-depth. Q-Synth and TB-OLSQ2 get (CX)-depths that are less than 10\% off from the optimal depths. Q-Synth (including Q-S+Qui) guarantees optimal CX-count. \quills{} (optimizing for depth) is on average 9\% off from the CX-count optimum (for N=129). It performs the same on CX-count as OLSQ2 (for N=93). The other tool combinations have 1-4\% worse CX-count than Q-Synth, except the heuristic tool SABRE (15\% off).

We conclude that \quills{} finds 5-9\% better depths and CX-depths on average than the optimal tools with different optimization goals, while finding 46\% better (CX)-depths on average than SABRE. This comes at the cost of larger CX-counts than other tools.
We observe that first optimizing SWAP count and then depth produces very good results overall.

\subsection{Impact of Optimization Metrics on Noise Reduction}
\label{sec:noise-sim}

Our ultimate motivation for circuit optimization is noise reduction.
The noise in a quantum circuit comes from noisy gates, interference from the environment, and measurement errors. One way to evaluate the impact of these errors is by running the same circuit many times and comparing the distribution of the measurement outcomes to the expected distribution. These distributions can be obtained, for instance by Qiskit's simulator with and without noise models.

Each quantum platform in the Qiskit library comes with a noise model, providing noise probabilities for native gates.
To use these, we first transpile our circuits to IBM's native gates, using Qiskit (without optimization).

The noise profile for gates varies significantly between the different qubits of the platform. This poses a problem for us: The same synthesized circuit with different initial mappings would have very different levels of noise.
Noise models are represented as a set of probability distributions over quantum operations, so we created an averaged version of each noise model by averaging the noise distribution per operation over all qubits. In this way, the noise model is still realistic, but does not depend on the initial mapping.

Qiskit's noise models specify gate errors and measurement errors, but not the interference from the environment when qubits are idle (also called decoherence). To incorporate this, we use the same approach as in~\cite{id-gates}: Qiskit provides a noise distribution for the \texttt{I}-gate, which is meant to model decoherence~\cite{qiskit-i-gate}. Hence, before simulating a circuit, we insert I-gates for all idle positions in that circuit. 

We apply all the different QLS tools (with the variations discussed in earlier experiments) to the transpiled circuits,
and simulate the mapped circuits under the averaged noise model.

\paragraph*{Experimental setup}
We simulate each synthesized circuit 100,000 times with and without noise. When simulating a circuit, we measure the mapped qubits at the very end of the circuit to obtain a probability distribution. We use Qiskit to compute the Hellinger distance between the two distributions obtained from simulations with and without the noise model. This is standard in the literature~\cite{noise1, noise2, noise3}.
Given the resource intensity of simulation, we only use the medium-sized platform IBM Guadalupe (16q) with the standard and VQE benchmarks. 

\begin{figure}[h]
    \centering
    \begin{subfigure}[t]{0.24\textwidth}
        \includegraphics[trim={3px 3px 3px 3px},clip,width=\linewidth]{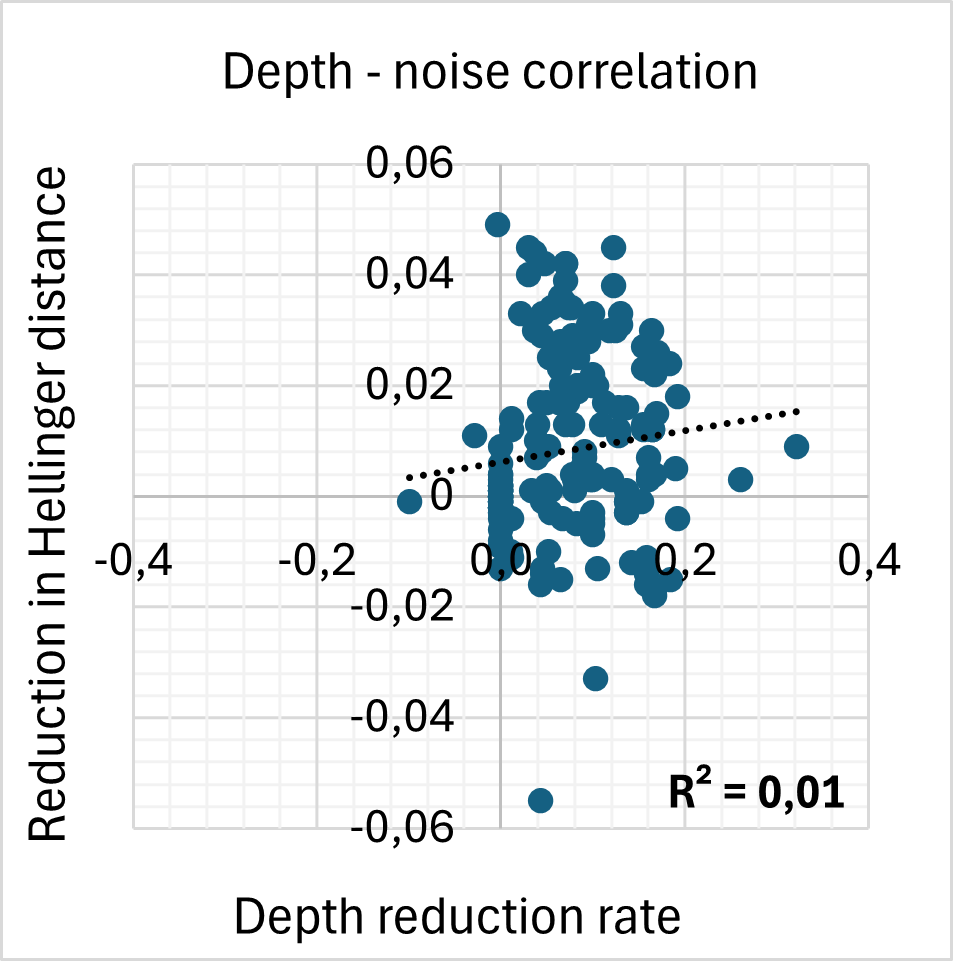}
        \label{fig:depth-noise}
        \caption{
        Noise coefficient 1\%.}
    \end{subfigure}
    \begin{subfigure}[t]{0.24\textwidth}
        \includegraphics[trim={3px 3px 3px 3px},clip,width=\linewidth]{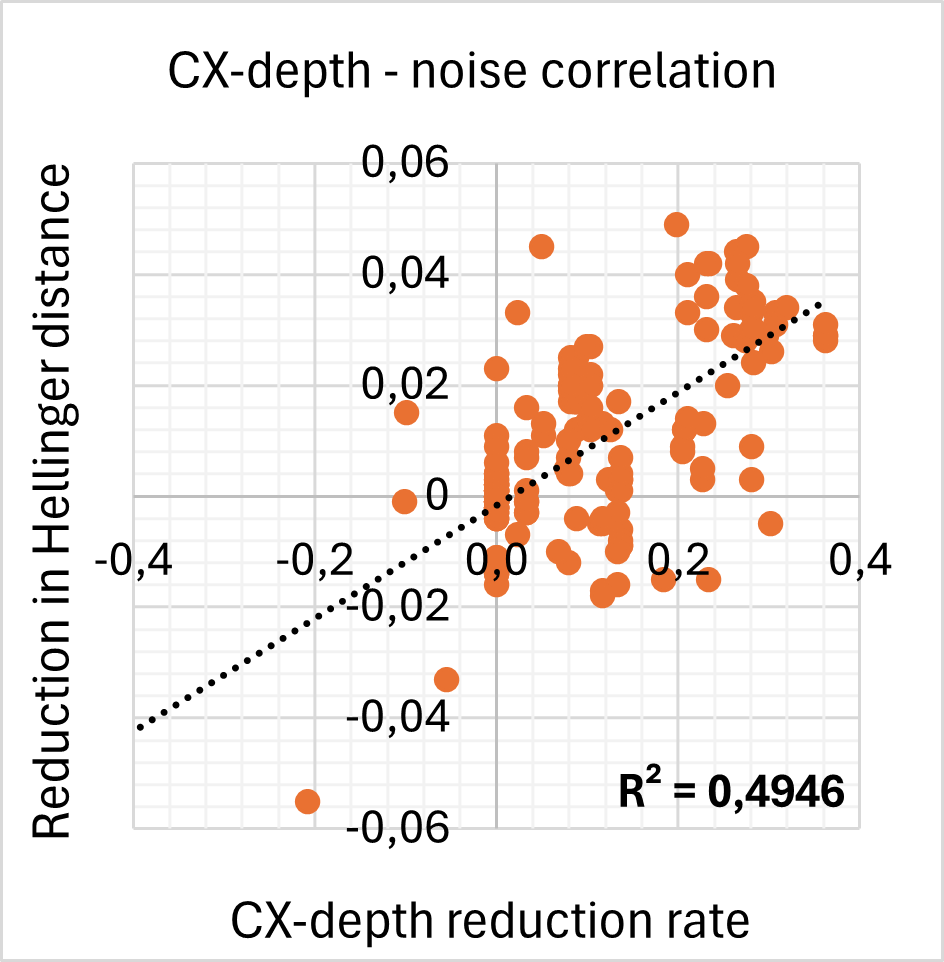}
        \label{fig:cx-depth-noise}
        \caption{
        Noise coefficient 49\%.}
    \end{subfigure}
    \begin{subfigure}[t]{0.24\textwidth}
        \includegraphics[trim={3px 3px 3px 3px},clip,width=\linewidth]{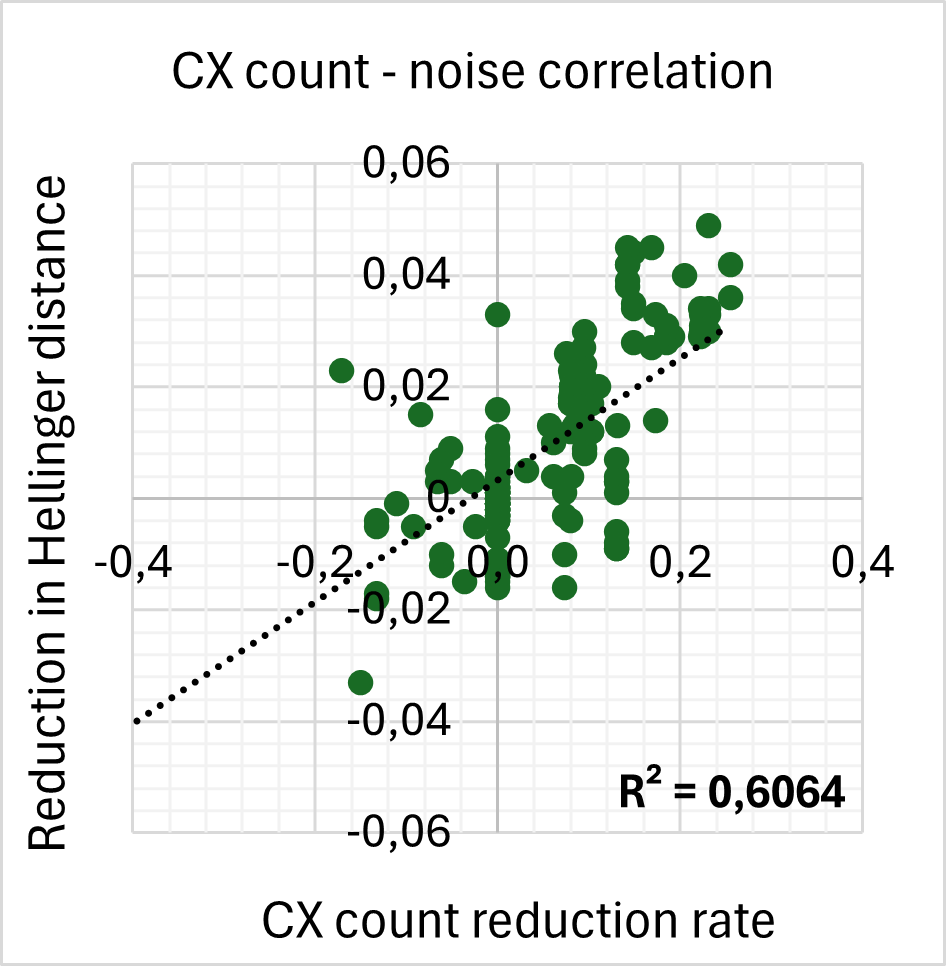}
        \label{fig:cx-count-noise}
        \caption{
        Noise coefficient 61\%.}
    \end{subfigure}
    \begin{subfigure}[t]{0.24\textwidth}
        \includegraphics[trim={3px 3px 3px 3px},clip,width=\linewidth]{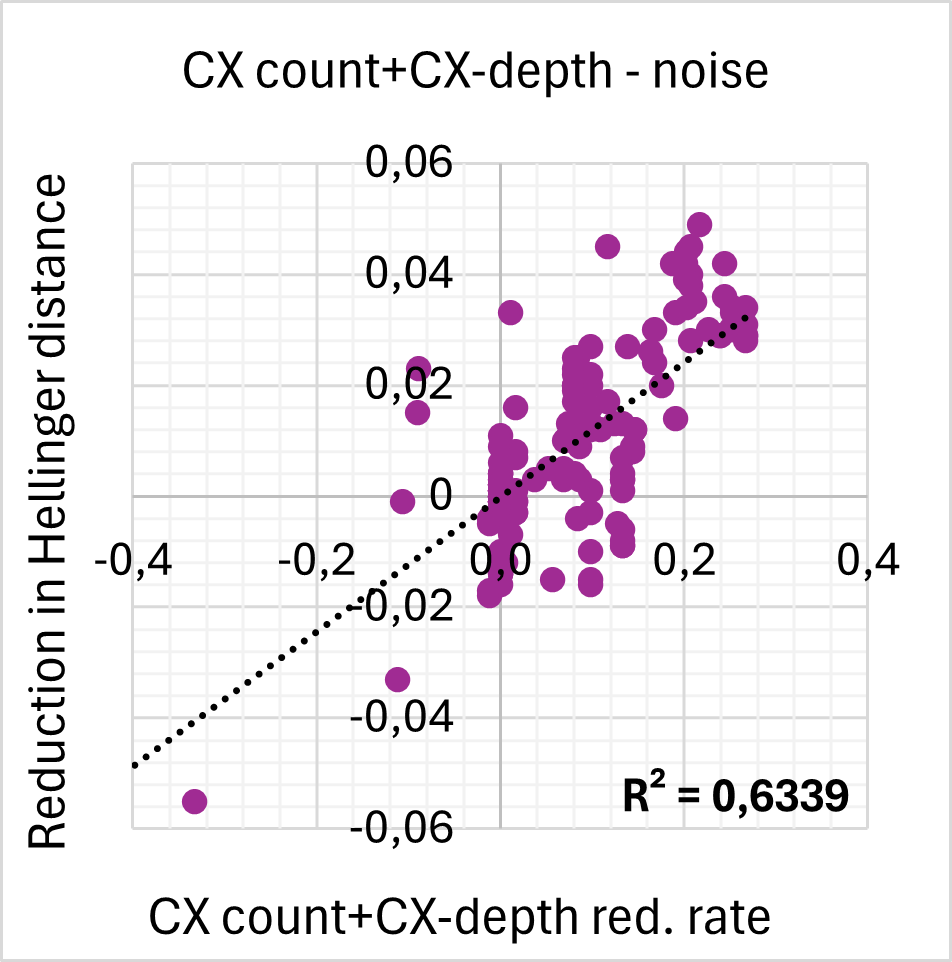}
        \label{fig:cx-count-cx-depth-noise}
        \caption{
        Noise coefficient 63\%.}
    \end{subfigure}
    \caption{Noise correlation coefficients between reduction rate for the optimization goal and reduction in Hellinger distance.
    }
    \label{fig:noise-correlations}
\end{figure}

\paragraph*{Results}
Figure \ref{fig:noise-correlations} shows the correlations between reduction rates for depth, CX-depth, CX count, and the sum of CX-depth and CX count and the reduction in the Hellinger distance as a measure for noise. We consider SABRE as baseline and report reductions in various tools. The depth reduction rate for tool X is normalized as $(d(\textit{SABRE})-d(X))/d(\textit{SABRE})$, and similar for the other metrics.
This presentation and analysis style follows the example of~\cite{resynth-noise}. 

\paragraph*{Discussion}
Figure \ref{fig:noise-correlations} shows that the depth reduction rate is not correlated with noise reduction at all, while the CX-depth reduction rate and the CX gate count reduction rate are quite strongly correlated.
Surprisingly, CX gate count reduction correlates more with noise than CX-depth reduction.
The correlation is strengthened further when looking at the sum of CX-depth and CX gate count. 

This result is surprising since it is generally assumed that noise scales with depth~\cite{resynth-noise, id-gates, sqgm, olsq}. For example, the authors of~\cite{sqgm} stress the importance of considering unary gates when minimizing depth. In~\cite{tket}, it is argued that minimizing CX-depth provides a better noise reduction than minimizing depth, since the binary CX gates are more noisy.
But seeing no correlation between depth and noise is unexpected.

We conclude that reducing CX gate count and CX-depth reduces noise in output circuits.
Further, CX gate count correlates better than CX-depth.
CX count + CX-depth provides the best correlation so far.
Surprisingly, we see no correlation between reducing depth and reducing noise.

\section{Conclusion}
In this paper, we have presented an efficient SAT encoding of (CX)-depth-optimal quantum layout synthesis. The SAT encoding has variations that allow ancillary SWAPs and finding locally minimal SWAP counts. We have implemented these encodings in the tool \quills{} and tested them on standard benchmarks. We compare \quills{} to the state-of-the-art depth-optimal tool, OLSQ2~\cite{olsq}, its near-optimal variant TB-OLSQ2, the state-of-the-art size optimal tool, Q-Synth~\cite{qsynth_sat}, and a heuristic, fast but sub-optimal tool, SABRE~\cite{sabre}. \quills{} and OLSQ2 synthesize circuits with better (CX)-depths than the other tools, and \quills{} vastly outperforms the previous best depth-optimal tool OLSQ2, by solving more instances, and often being more than 10-100x faster. 

We also evaluate the amount of noise in the circuits synthesized by each tool, by measuring the Hellinger distance between ideal and noisy simulations of the circuits. We see that reducing the number of SWAPs and reducing CX-depth (and in particular their combination) is strongly correlated with reducing noise. Contrary to what is generally assumed in the literature, we do not see a correlation between reducing depth and reducing noise. However, this should be explored further, by testing whether these results carry over to larger quantum platforms and when executing the circuits on real quantum computers.

\bibliography{refs}

\newpage

\appendix

\section{Benchmarks Information}
\label{app:Benchmarks}
The circuit data for all the circuits used in the experiments can be found in Table \ref{fig:circuit-data}.

\begin{table}[H]
\caption{Circuit data for all standard, VQE and QUEKO benchmarks.}
\label{fig:circuit-data}
\footnotesize
\centering
\begin{tabular}{|l|lllll|}
\hline
\textbf{Circuit name}           & \textbf{Qubits} & \textbf{Gates} & \textbf{CX gates} & \textbf{Depth} & \textbf{CX-depth} \\ \hline
4gt13\_92                  & 5               & 66             & 30                & 38             & 26                \\
4mod5-v1\_22               & 5               & 21             & 11                & 12             & 10                \\
adder                      & 4               & 23             & 10                & 11             & 6                 \\
barenco\_tof\_4            & 7               & 72             & 34                & 68             & 34                \\
barenco\_tof\_5            & 9               & 104            & 50                & 95             & 48                \\
mod\_mult\_55              & 9               & 91             & 40                & 47             & 25                \\
mod5mils\_65               & 5               & 35             & 16                & 21             & 16                \\
or                         & 3               & 17             & 6                 & 8              & 6                 \\
qaoa5                      & 5               & 22             & 8                 & 14             & 8                 \\
qft\_8                     & 8               & 106            & 56                & 42             & 26                \\
rc\_adder\_6               & 14              & 140            & 71                & 83             & 40                \\
tof\_4                     & 7               & 55             & 22                & 46             & 21                \\
tof\_5                     & 9               & 75             & 30                & 61             & 28                \\
toffoli                    & 3               & 15             & 6                 & 11             & 6                 \\
vbe\_adder\_3              & 10              & 89             & 50                & 58             & 33                \\ \hline
vqe\_8\_0\_5\_100          & 8               & 96             & 52                & 79             & 44                \\
vqe\_8\_0\_10\_100     & 8               & 129            & 63                & 92             & 51                \\
vqe\_8\_1\_5\_100      & 8               & 43             & 18                & 32             & 16                \\
vqe\_8\_1\_10\_100     & 8               & 100            & 47                & 76             & 41                \\
vqe\_8\_2\_5\_100      & 8               & 90             & 48                & 80             & 44                \\
vqe\_8\_2\_10\_100     & 8               & 153            & 79                & 136            & 75                \\
vqe\_8\_3\_5\_100      & 8               & 76             & 40                & 61             & 35                \\
vqe\_8\_3\_10\_100     & 8               & 151            & 78                & 119            & 68                \\
vqe\_8\_4\_5\_100      & 8               & 76             & 39                & 60             & 32                \\
vqe\_8\_4\_10\_100     & 8               & 136            & 71                & 102            & 56                \\ \hline
16QBT\_05CYC\_TFL\_0 & 16              & 37             & 15                & 5              & 5                 \\
16QBT\_10CYC\_TFL\_0 & 16              & 73             & 29                & 10             & 7                 \\
16QBT\_15CYC\_TFL\_0 & 16              & 109            & 44                & 15             & 11                \\
16QBT\_20CYC\_TFL\_0 & 16              & 145            & 58                & 20             & 14                \\
16QBT\_25CYC\_TFL\_0 & 16              & 180            & 72                & 25             & 15                \\
16QBT\_30CYC\_TFL\_0 & 16              & 217            & 87                & 30             & 18                \\
16QBT\_35CYC\_TFL\_0 & 16              & 253            & 101               & 35             & 25                \\
16QBT\_40CYC\_TFL\_0 & 16              & 289            & 116               & 40             & 27                \\
16QBT\_45CYC\_TFL\_0 & 16              & 325            & 130               & 45             & 30                \\
54QBT\_05CYC\_QSE\_0 & 54              & 192            & 54                & 5              & 5                 \\
54QBT\_10CYC\_QSE\_0 & 54              & 384            & 108               & 10             & 10                \\
54QBT\_15CYC\_QSE\_0 & 54              & 576            & 162               & 15             & 12                \\
54QBT\_20CYC\_QSE\_0 & 54              & 767            & 216               & 20             & 16                \\
54QBT\_25CYC\_QSE\_0 & 54              & 959            & 270               & 25             & 21                \\
54QBT\_30CYC\_QSE\_0 & 54              & 1151           & 324               & 30             & 24                \\
54QBT\_35CYC\_QSE\_0 & 54              & 1342           & 378               & 35             & 32                \\
54QBT\_40CYC\_QSE\_0 & 54              & 1534           & 432               & 40             & 33                \\
54QBT\_45CYC\_QSE\_0 & 54              & 1727           & 487               & 45             & 38                \\ \hline
\end{tabular}
\end{table}

\newpage

\section{Runtime data}
\label{app:runtimedata}

The following are tables that show run-time performance for all tools. Note that the tables are separated by platform and \quills{}, TB-OLSQ2 and OLSQ2 are synthesizing depth-optimal circuits with locally minimal SWAP counts.

\begin{table}[H]
\caption{Run time for all tools on the Tenerife platform.}
\label{fig:tenerife-time}
\centering
\begin{tabular}{|l|lllll|}
\hline
\multicolumn{1}{|c|}{Tenerife} & \multicolumn{5}{c|}{Run time (s)}                                                        \\ \hline
Circuit                            & \multicolumn{1}{l|}{SABRE} & \multicolumn{1}{l|}{Q-Synth} & \multicolumn{1}{l|}{QuilLS} & \multicolumn{1}{l|}{TB-OLSQ2} & OLSQ2  \\ \hline
4gt13\_92                          & \multicolumn{1}{l|}{0.003} & \multicolumn{1}{l|}{0.283}   & \multicolumn{1}{l|}{0.596}  & \multicolumn{1}{l|}{1.783}    & 95.054 \\
4mod5-v1\_22                       & \multicolumn{1}{l|}{0.002} & \multicolumn{1}{l|}{0.067}   & \multicolumn{1}{l|}{0.096}  & \multicolumn{1}{l|}{0.923}    & 12.01  \\
adder                              & \multicolumn{1}{l|}{0.002} & \multicolumn{1}{l|}{0.065}   & \multicolumn{1}{l|}{0.126}  & \multicolumn{1}{l|}{0.827}    & 10.93  \\
mod5mils\_65                       & \multicolumn{1}{l|}{0.003} & \multicolumn{1}{l|}{0.088}   & \multicolumn{1}{l|}{0.299}  & \multicolumn{1}{l|}{1.485}    & 36.403 \\
or                                 & \multicolumn{1}{l|}{0.002} & \multicolumn{1}{l|}{0.056}   & \multicolumn{1}{l|}{0.023}  & \multicolumn{1}{l|}{0.612}    & 4.758  \\
qaoa5                              & \multicolumn{1}{l|}{0.002} & \multicolumn{1}{l|}{0.062}   & \multicolumn{1}{l|}{0.055}  & \multicolumn{1}{l|}{0.638}    & 11.625 \\
toffoli                            & \multicolumn{1}{l|}{0.002} & \multicolumn{1}{l|}{0.056}   & \multicolumn{1}{l|}{0.029}  & \multicolumn{1}{l|}{0.596}    & 6.04   \\ \hline
\end{tabular}
\end{table}

\begin{table}[H]
\caption{Run time for all tools on the Melbourne platform.}
\label{fig:melbourne-time}
\centering
\begin{tabular}{|l|lllll|}
\hline
\multicolumn{1}{|c|}{Melbourne} & \multicolumn{5}{c|}{Run time (s)}                                                                                                     \\ \hline
Circuit                         & \multicolumn{1}{l|}{SABRE} & \multicolumn{1}{l|}{Q-Synth}  & \multicolumn{1}{l|}{QuilLS}   & \multicolumn{1}{l|}{TB-OLSQ2} & OLSQ2    \\ \hline
4gt13\_92                       & \multicolumn{1}{l|}{0.006} & \multicolumn{1}{l|}{1.301}    & \multicolumn{1}{l|}{68.036}   & \multicolumn{1}{l|}{96.982}   & 4512.505 \\
4mod5-v1\_22                    & \multicolumn{1}{l|}{0.003} & \multicolumn{1}{l|}{0.128}    & \multicolumn{1}{l|}{2.105}    & \multicolumn{1}{l|}{3.103}    & 163.763  \\
adder                           & \multicolumn{1}{l|}{0.007} & \multicolumn{1}{l|}{0.067}    & \multicolumn{1}{l|}{0.13}     & \multicolumn{1}{l|}{1.887}    & 26.116   \\
barenco\_tof\_4                 & \multicolumn{1}{l|}{0.005} & \multicolumn{1}{l|}{0.999}    & \multicolumn{1}{l|}{23.11}    & \multicolumn{1}{l|}{80.469}   & 1186.621 \\
barenco\_tof\_5                 & \multicolumn{1}{l|}{0.007} & \multicolumn{1}{l|}{1.206}    & \multicolumn{1}{l|}{99.803}   & \multicolumn{1}{l|}{167.255}  & TO     \\
mod\_mult\_55                   & \multicolumn{1}{l|}{0.007} & \multicolumn{1}{l|}{15.928}   & \multicolumn{1}{l|}{62.827}   & \multicolumn{1}{l|}{251.428}  & 4428.399 \\
mod5mils\_65                    & \multicolumn{1}{l|}{0.004} & \multicolumn{1}{l|}{0.355}    & \multicolumn{1}{l|}{12.251}   & \multicolumn{1}{l|}{25.87}    & 568.597  \\
or                              & \multicolumn{1}{l|}{0.003} & \multicolumn{1}{l|}{0.094}    & \multicolumn{1}{l|}{1.058}    & \multicolumn{1}{l|}{2.156}    & 58.896   \\
qaoa5                           & \multicolumn{1}{l|}{0.003} & \multicolumn{1}{l|}{0.07}     & \multicolumn{1}{l|}{0.167}    & \multicolumn{1}{l|}{1.91}     & 33.524   \\
qft\_8                          & \multicolumn{1}{l|}{0.007} & \multicolumn{1}{l|}{4467.153} & \multicolumn{1}{l|}{356.405}  & \multicolumn{1}{l|}{139.898}  & TO     \\
rc\_adder\_6                    & \multicolumn{1}{l|}{0.009} & \multicolumn{1}{l|}{256.357}  & \multicolumn{1}{l|}{TO}     & \multicolumn{1}{l|}{105.723}  & TO     \\
tof\_4                          & \multicolumn{1}{l|}{0.004} & \multicolumn{1}{l|}{0.117}    & \multicolumn{1}{l|}{6.686}    & \multicolumn{1}{l|}{4.049}    & 710.947  \\
tof\_5                          & \multicolumn{1}{l|}{0.005} & \multicolumn{1}{l|}{0.141}    & \multicolumn{1}{l|}{21.759}   & \multicolumn{1}{l|}{5.132}    & 1367.514 \\
toffoli                         & \multicolumn{1}{l|}{0.003} & \multicolumn{1}{l|}{0.077}    & \multicolumn{1}{l|}{0.16}     & \multicolumn{1}{l|}{1.477}    & 42.567   \\
vbe\_adder\_3                   & \multicolumn{1}{l|}{0.006} & \multicolumn{1}{l|}{8.292}    & \multicolumn{1}{l|}{477.227}  & \multicolumn{1}{l|}{110.356}  & TO     \\
vqe\_8\_0\_5\_100               & \multicolumn{1}{l|}{0.007} & \multicolumn{1}{l|}{37.794}   & \multicolumn{1}{l|}{54.959}   & \multicolumn{1}{l|}{256.087}  & TO     \\
vqe\_8\_0\_10\_100              & \multicolumn{1}{l|}{0.011} & \multicolumn{1}{l|}{50.872}   & \multicolumn{1}{l|}{244.716}  & \multicolumn{1}{l|}{310.685}  & TO     \\
vqe\_8\_1\_5\_100               & \multicolumn{1}{l|}{0.004} & \multicolumn{1}{l|}{0.279}    & \multicolumn{1}{l|}{1.023}    & \multicolumn{1}{l|}{4.873}    & 409.802  \\
vqe\_8\_1\_10\_100              & \multicolumn{1}{l|}{0.007} & \multicolumn{1}{l|}{12.795}   & \multicolumn{1}{l|}{145.188}  & \multicolumn{1}{l|}{208.231}  & TO     \\
vqe\_8\_2\_5\_100               & \multicolumn{1}{l|}{0.007} & \multicolumn{1}{l|}{5.02}     & \multicolumn{1}{l|}{44.332}   & \multicolumn{1}{l|}{164.869}  & TO     \\
vqe\_8\_2\_10\_100              & \multicolumn{1}{l|}{0.009} & \multicolumn{1}{l|}{97.097}   & \multicolumn{1}{l|}{132.278}  & \multicolumn{1}{l|}{1312.836} & TO     \\
vqe\_8\_3\_5\_100               & \multicolumn{1}{l|}{0.006} & \multicolumn{1}{l|}{5.525}    & \multicolumn{1}{l|}{43.39}    & \multicolumn{1}{l|}{132.728}  & 2935.913 \\
vqe\_8\_3\_10\_100              & \multicolumn{1}{l|}{0.012} & \multicolumn{1}{l|}{399.071}  & \multicolumn{1}{l|}{1051.673} & \multicolumn{1}{l|}{423.518}  & TO     \\
vqe\_8\_4\_5\_100               & \multicolumn{1}{l|}{0.007} & \multicolumn{1}{l|}{4.96}     & \multicolumn{1}{l|}{29.369}   & \multicolumn{1}{l|}{31.418}   & 1473.159 \\
vqe\_8\_4\_10\_100              & \multicolumn{1}{l|}{0.009} & \multicolumn{1}{l|}{92.303}   & \multicolumn{1}{l|}{253.466}  & \multicolumn{1}{l|}{635.507}  & TO     \\ \hline
\end{tabular}
\end{table}

\begin{table}[H]
\caption{Run time for all tools on the Guadalupe platform.}
\label{fig:guadalupe-time}
\centering
\begin{tabular}{|l|lllll|}
\hline
\multicolumn{1}{|c|}{Guadalupe} & \multicolumn{5}{c|}{Run time (s)}                                                                                                     \\ \hline
Circuit                         & \multicolumn{1}{l|}{SABRE} & \multicolumn{1}{l|}{Q-Synth}  & \multicolumn{1}{l|}{QuilLS}   & \multicolumn{1}{l|}{TB-OLSQ2} & OLSQ2    \\ \hline
4gt13\_92                       & \multicolumn{1}{l|}{0.008} & \multicolumn{1}{l|}{8.747}    & \multicolumn{1}{l|}{550.423}  & \multicolumn{1}{l|}{248.423}  & TO     \\
4mod5-v1\_22                    & \multicolumn{1}{l|}{0.004} & \multicolumn{1}{l|}{0.169}    & \multicolumn{1}{l|}{2.551}    & \multicolumn{1}{l|}{4.846}    & 663.422  \\
adder                           & \multicolumn{1}{l|}{0.004} & \multicolumn{1}{l|}{0.193}    & \multicolumn{1}{l|}{1.467}    & \multicolumn{1}{l|}{3.725}    & 349.219  \\
barenco\_tof\_4                 & \multicolumn{1}{l|}{0.007} & \multicolumn{1}{l|}{4.523}    & \multicolumn{1}{l|}{25.454}   & \multicolumn{1}{l|}{192.285}  & TO     \\
barenco\_tof\_5                 & \multicolumn{1}{l|}{0.008} & \multicolumn{1}{l|}{20.845}   & \multicolumn{1}{l|}{98.344}   & \multicolumn{1}{l|}{726.629}  & TO     \\
mod\_mult\_55                   & \multicolumn{1}{l|}{0.008} & \multicolumn{1}{l|}{130.651}  & \multicolumn{1}{l|}{516.587}  & \multicolumn{1}{l|}{671.409}  & TO     \\
mod5mils\_65                    & \multicolumn{1}{l|}{0.004} & \multicolumn{1}{l|}{0.621}    & \multicolumn{1}{l|}{16.434}   & \multicolumn{1}{l|}{40.996}   & 3950.593 \\
or                              & \multicolumn{1}{l|}{0.004} & \multicolumn{1}{l|}{0.101}    & \multicolumn{1}{l|}{0.815}    & \multicolumn{1}{l|}{2.98}     & 466.202  \\
qaoa5                           & \multicolumn{1}{l|}{0.004} & \multicolumn{1}{l|}{0.076}    & \multicolumn{1}{l|}{0.181}    & \multicolumn{1}{l|}{3.279}    & 303.088  \\
qft\_8                          & \multicolumn{1}{l|}{0.008} & \multicolumn{1}{l|}{2799.019} & \multicolumn{1}{l|}{2201.408} & \multicolumn{1}{l|}{498.312}  & TO     \\
rc\_adder\_6                    & \multicolumn{1}{l|}{0.011} & \multicolumn{1}{l|}{TO}     & \multicolumn{1}{l|}{1397.478} & \multicolumn{1}{l|}{1275.453} & TO     \\
tof\_4                          & \multicolumn{1}{l|}{0.005} & \multicolumn{1}{l|}{0.312}    & \multicolumn{1}{l|}{8.162}    & \multicolumn{1}{l|}{8.344}    & 5607.296 \\
tof\_5                          & \multicolumn{1}{l|}{0.006} & \multicolumn{1}{l|}{1.579}    & \multicolumn{1}{l|}{62.628}   & \multicolumn{1}{l|}{139.053}  & TO     \\
toffoli                         & \multicolumn{1}{l|}{0.003} & \multicolumn{1}{l|}{0.08}     & \multicolumn{1}{l|}{0.257}    & \multicolumn{1}{l|}{2.575}    & 362.603  \\
vbe\_adder\_3                   & \multicolumn{1}{l|}{0.007} & \multicolumn{1}{l|}{12.906}   & \multicolumn{1}{l|}{136.853}  & \multicolumn{1}{l|}{204.682}  & TO     \\
vqe\_8\_0\_5\_100               & \multicolumn{1}{l|}{0.009} & \multicolumn{1}{l|}{775.064}  & \multicolumn{1}{l|}{636.835}  & \multicolumn{1}{l|}{611.845}  & TO     \\
vqe\_8\_0\_10\_100              & \multicolumn{1}{l|}{0.01}  & \multicolumn{1}{l|}{372.767}  & \multicolumn{1}{l|}{3147.816} & \multicolumn{1}{l|}{1632.732} & TO     \\
vqe\_8\_1\_5\_100               & \multicolumn{1}{l|}{0.004} & \multicolumn{1}{l|}{0.261}    & \multicolumn{1}{l|}{8.348}    & \multicolumn{1}{l|}{7.385}    & 3085.1   \\
vqe\_8\_1\_10\_100              & \multicolumn{1}{l|}{0.008} & \multicolumn{1}{l|}{56.059}   & \multicolumn{1}{l|}{280.444}  & \multicolumn{1}{l|}{400.851}  & TO     \\
vqe\_8\_2\_5\_100               & \multicolumn{1}{l|}{0.007} & \multicolumn{1}{l|}{19.29}    & \multicolumn{1}{l|}{96.299}   & \multicolumn{1}{l|}{343.231}  & TO     \\
vqe\_8\_2\_10\_100              & \multicolumn{1}{l|}{0.012} & \multicolumn{1}{l|}{1200.897} & \multicolumn{1}{l|}{886.175}  & \multicolumn{1}{l|}{2680.416} & TO     \\
vqe\_8\_3\_5\_100               & \multicolumn{1}{l|}{0.007} & \multicolumn{1}{l|}{11.135}   & \multicolumn{1}{l|}{515.506}  & \multicolumn{1}{l|}{339.764}  & TO     \\
vqe\_8\_3\_10\_100              & \multicolumn{1}{l|}{0.012} & \multicolumn{1}{l|}{TO}     & \multicolumn{1}{l|}{3931.779} & \multicolumn{1}{l|}{6208.931} & TO     \\
vqe\_8\_4\_5\_100               & \multicolumn{1}{l|}{0.006} & \multicolumn{1}{l|}{5.835}    & \multicolumn{1}{l|}{150.53}   & \multicolumn{1}{l|}{185.274}  & TO     \\
vqe\_8\_4\_10\_100              & \multicolumn{1}{l|}{0.01}  & \multicolumn{1}{l|}{6609.111} & \multicolumn{1}{l|}{6817.128} & \multicolumn{1}{l|}{3642.612} & TO     \\
16QBT\_05CYC\_TFL\_0            & \multicolumn{1}{l|}{0.004} & \multicolumn{1}{l|}{0.302}    & \multicolumn{1}{l|}{0.136}    & \multicolumn{1}{l|}{3.577}    & 15.014   \\
16QBT\_10CYC\_TFL\_0            & \multicolumn{1}{l|}{0.011} & \multicolumn{1}{l|}{2.369}    & \multicolumn{1}{l|}{5.697}    & \multicolumn{1}{l|}{5.525}    & 522.863  \\
16QBT\_15CYC\_TFL\_0            & \multicolumn{1}{l|}{0.023} & \multicolumn{1}{l|}{6283.461} & \multicolumn{1}{l|}{1061.536} & \multicolumn{1}{l|}{153.298}  & TO     \\
16QBT\_20CYC\_TFL\_0            & \multicolumn{1}{l|}{0.039} & \multicolumn{1}{l|}{TO}     & \multicolumn{1}{l|}{TO}     & \multicolumn{1}{l|}{1549.09}  & TO     \\
16QBT\_25CYC\_TFL\_0            & \multicolumn{1}{l|}{0.088} & \multicolumn{1}{l|}{TO}     & \multicolumn{1}{l|}{TO}     & \multicolumn{1}{l|}{TO}     & TO     \\
16QBT\_30CYC\_TFL\_0            & \multicolumn{1}{l|}{0.078} & \multicolumn{1}{l|}{TO}     & \multicolumn{1}{l|}{TO}     & \multicolumn{1}{l|}{5748.028} & TO     \\
16QBT\_35CYC\_TFL\_0            & \multicolumn{1}{l|}{0.11}  & \multicolumn{1}{l|}{TO}     & \multicolumn{1}{l|}{1758.855} & \multicolumn{1}{l|}{1679.068} & TO     \\
16QBT\_40CYC\_TFL\_0            & \multicolumn{1}{l|}{0.099} & \multicolumn{1}{l|}{TO}     & \multicolumn{1}{l|}{TO}     & \multicolumn{1}{l|}{TO}     & TO     \\
16QBT\_45CYC\_TFL\_0            & \multicolumn{1}{l|}{0.077} & \multicolumn{1}{l|}{TO}     & \multicolumn{1}{l|}{TO}     & \multicolumn{1}{l|}{TO}     & TO     \\ \hline
\end{tabular}
\end{table}

\begin{table}[H]
\caption{Run time for all tools on the Tokyo platform.}
\label{fig:tokyo-time}
\centering
\begin{tabular}{|l|lllll|}
\hline
\multicolumn{1}{|c|}{Tokyo} & \multicolumn{5}{c|}{Run time (s)}                                                                                                    \\ \hline
Circuit                     & \multicolumn{1}{l|}{SABRE} & \multicolumn{1}{l|}{Q-Synth} & \multicolumn{1}{l|}{QuilLS}   & \multicolumn{1}{l|}{TB-OLSQ2} & OLSQ2    \\ \hline
4gt13\_92                   & \multicolumn{1}{l|}{0.006} & \multicolumn{1}{l|}{0.147}   & \multicolumn{1}{l|}{2.794}    & \multicolumn{1}{l|}{7.385}    & 520.98   \\
4mod5-v1\_22                & \multicolumn{1}{l|}{0.004} & \multicolumn{1}{l|}{0.039}   & \multicolumn{1}{l|}{0.292}    & \multicolumn{1}{l|}{4.137}    & 60.501   \\
adder                       & \multicolumn{1}{l|}{0.003} & \multicolumn{1}{l|}{0.037}   & \multicolumn{1}{l|}{0.23}     & \multicolumn{1}{l|}{3.766}    & 53.443   \\
barenco\_tof\_4             & \multicolumn{1}{l|}{0.005} & \multicolumn{1}{l|}{0.058}   & \multicolumn{1}{l|}{4.966}    & \multicolumn{1}{l|}{8.774}    & 1067.887 \\
barenco\_tof\_5             & \multicolumn{1}{l|}{0.008} & \multicolumn{1}{l|}{0.072}   & \multicolumn{1}{l|}{37.271}   & \multicolumn{1}{l|}{11.531}   & 2226.538 \\
mod\_mult\_55               & \multicolumn{1}{l|}{0.006} & \multicolumn{1}{l|}{2.549}   & \multicolumn{1}{l|}{31.621}   & \multicolumn{1}{l|}{38.295}   & 4556.548 \\
mod5mils\_65                & \multicolumn{1}{l|}{0.004} & \multicolumn{1}{l|}{0.042}   & \multicolumn{1}{l|}{0.704}    & \multicolumn{1}{l|}{5.159}    & 159.663  \\
or                          & \multicolumn{1}{l|}{0.004} & \multicolumn{1}{l|}{0.034}   & \multicolumn{1}{l|}{0.111}    & \multicolumn{1}{l|}{2.861}    & 29.848   \\
qaoa5                       & \multicolumn{1}{l|}{0.004} & \multicolumn{1}{l|}{0.039}   & \multicolumn{1}{l|}{0.292}    & \multicolumn{1}{l|}{3.681}    & 69.902   \\
qft\_8                      & \multicolumn{1}{l|}{0.008} & \multicolumn{1}{l|}{21.863}  & \multicolumn{1}{l|}{2992.915} & \multicolumn{1}{l|}{2077.731} & TO     \\
rc\_adder\_6                & \multicolumn{1}{l|}{0.01}  & \multicolumn{1}{l|}{4.061}   & \multicolumn{1}{l|}{1158.291} & \multicolumn{1}{l|}{60.531}   & TO     \\
tof\_4                      & \multicolumn{1}{l|}{0.004} & \multicolumn{1}{l|}{0.05}    & \multicolumn{1}{l|}{2.785}    & \multicolumn{1}{l|}{6.493}    & 552.836  \\
tof\_5                      & \multicolumn{1}{l|}{0.005} & \multicolumn{1}{l|}{0.061}   & \multicolumn{1}{l|}{4.974}    & \multicolumn{1}{l|}{8.29}     & 954.309  \\
toffoli                     & \multicolumn{1}{l|}{0.003} & \multicolumn{1}{l|}{0.034}   & \multicolumn{1}{l|}{0.152}    & \multicolumn{1}{l|}{2.831}    & 38.516   \\
vbe\_adder\_3               & \multicolumn{1}{l|}{0.006} & \multicolumn{1}{l|}{0.07}    & \multicolumn{1}{l|}{26.584}   & \multicolumn{1}{l|}{13.495}   & 1875.32  \\
vqe\_8\_0\_5\_100           & \multicolumn{1}{l|}{0.01}  & \multicolumn{1}{l|}{1.839}   & \multicolumn{1}{l|}{320.795}  & \multicolumn{1}{l|}{47.34}    & TO     \\
vqe\_8\_0\_10\_100          & \multicolumn{1}{l|}{0.008} & \multicolumn{1}{l|}{10.14}   & \multicolumn{1}{l|}{324.179}  & \multicolumn{1}{l|}{118.798}  & TO     \\
vqe\_8\_1\_5\_100           & \multicolumn{1}{l|}{0.004} & \multicolumn{1}{l|}{0.052}   & \multicolumn{1}{l|}{1.617}    & \multicolumn{1}{l|}{6.213}    & 300.351  \\
vqe\_8\_1\_10\_100          & \multicolumn{1}{l|}{0.007} & \multicolumn{1}{l|}{4.278}   & \multicolumn{1}{l|}{139.346}  & \multicolumn{1}{l|}{66.397}   & TO     \\
vqe\_8\_2\_5\_100           & \multicolumn{1}{l|}{0.006} & \multicolumn{1}{l|}{0.351}   & \multicolumn{1}{l|}{59.182}   & \multicolumn{1}{l|}{13.572}   & 4940.026 \\
vqe\_8\_2\_10\_100          & \multicolumn{1}{l|}{0.01}  & \multicolumn{1}{l|}{3.58}    & \multicolumn{1}{l|}{1001.557} & \multicolumn{1}{l|}{56.413}   & TO     \\
vqe\_8\_3\_5\_100           & \multicolumn{1}{l|}{0.006} & \multicolumn{1}{l|}{0.342}   & \multicolumn{1}{l|}{57.057}   & \multicolumn{1}{l|}{28.309}   & 3272.329 \\
vqe\_8\_3\_10\_100          & \multicolumn{1}{l|}{0.012} & \multicolumn{1}{l|}{25.49}   & \multicolumn{1}{l|}{3493.999} & \multicolumn{1}{l|}{529.371}  & TO     \\
vqe\_8\_4\_5\_100           & \multicolumn{1}{l|}{0.006} & \multicolumn{1}{l|}{0.118}   & \multicolumn{1}{l|}{35.416}   & \multicolumn{1}{l|}{11.476}   & 2118.625 \\
vqe\_8\_4\_10\_100          & \multicolumn{1}{l|}{0.01}  & \multicolumn{1}{l|}{22.156}  & \multicolumn{1}{l|}{1432.25}  & \multicolumn{1}{l|}{79.98}    & TO     \\
16QBT\_05CYC\_TFL\_0        & \multicolumn{1}{l|}{0.003} & \multicolumn{1}{l|}{0.051}   & \multicolumn{1}{l|}{0.215}    & \multicolumn{1}{l|}{7.8}      & 33.684   \\
16QBT\_10CYC\_TFL\_0        & \multicolumn{1}{l|}{0.005} & \multicolumn{1}{l|}{0.065}   & \multicolumn{1}{l|}{1.394}    & \multicolumn{1}{l|}{10.575}   & 153.071  \\
16QBT\_15CYC\_TFL\_0        & \multicolumn{1}{l|}{0.007} & \multicolumn{1}{l|}{0.081}   & \multicolumn{1}{l|}{4.982}    & \multicolumn{1}{l|}{13.411}   & 342.658  \\
16QBT\_20CYC\_TFL\_0        & \multicolumn{1}{l|}{0.007} & \multicolumn{1}{l|}{0.105}   & \multicolumn{1}{l|}{21.897}   & \multicolumn{1}{l|}{17.8}     & 1016.615 \\
16QBT\_25CYC\_TFL\_0        & \multicolumn{1}{l|}{0.008} & \multicolumn{1}{l|}{0.23}    & \multicolumn{1}{l|}{123.929}  & \multicolumn{1}{l|}{23.09}    & TO     \\
16QBT\_30CYC\_TFL\_0        & \multicolumn{1}{l|}{0.01}  & \multicolumn{1}{l|}{0.105}   & \multicolumn{1}{l|}{41.819}   & \multicolumn{1}{l|}{20.688}   & 3127.651 \\
16QBT\_35CYC\_TFL\_0        & \multicolumn{1}{l|}{0.013} & \multicolumn{1}{l|}{0.114}   & \multicolumn{1}{l|}{94.534}   & \multicolumn{1}{l|}{22.436}   & 1963.847 \\
16QBT\_40CYC\_TFL\_0        & \multicolumn{1}{l|}{0.013} & \multicolumn{1}{l|}{19.464}  & \multicolumn{1}{l|}{1110.386} & \multicolumn{1}{l|}{451.63}   & TO     \\
16QBT\_45CYC\_TFL\_0        & \multicolumn{1}{l|}{0.014} & \multicolumn{1}{l|}{2.881}   & \multicolumn{1}{l|}{440.73}   & \multicolumn{1}{l|}{50.645}   & TO     \\ \hline
\end{tabular}
\end{table}

\begin{table}[H]
\caption{Run time for all tools on the Cambridge platform.}
\label{fig:cambridge-time}
\centering
\begin{tabular}{|l|lllll|}
\hline
\multicolumn{1}{|c|}{Cambridge} & \multicolumn{5}{c|}{Run time (s)}                                                                                                     \\ \hline
Circuit                         & \multicolumn{1}{l|}{SABRE} & \multicolumn{1}{l|}{Q-Synth}  & \multicolumn{1}{l|}{QuilLS}   & \multicolumn{1}{l|}{TB-OLSQ2} & OLSQ2    \\ \hline
4gt13\_92                       & \multicolumn{1}{l|}{0.008} & \multicolumn{1}{l|}{20.354}   & \multicolumn{1}{l|}{3671.575} & \multicolumn{1}{l|}{798.119}  & TO     \\
4mod5-v1\_22                    & \multicolumn{1}{l|}{0.005} & \multicolumn{1}{l|}{0.284}    & \multicolumn{1}{l|}{5.797}    & \multicolumn{1}{l|}{10.776}   & 6543.898 \\
adder                           & \multicolumn{1}{l|}{0.004} & \multicolumn{1}{l|}{0.334}    & \multicolumn{1}{l|}{6.735}    & \multicolumn{1}{l|}{6.785}    & 2924.108 \\
barenco\_tof\_4                 & \multicolumn{1}{l|}{0.007} & \multicolumn{1}{l|}{7.029}    & \multicolumn{1}{l|}{57.407}   & \multicolumn{1}{l|}{606.403}  & TO     \\
barenco\_tof\_5                 & \multicolumn{1}{l|}{0.009} & \multicolumn{1}{l|}{109.654}  & \multicolumn{1}{l|}{356.535}  & \multicolumn{1}{l|}{1646.73}  & TO     \\
mod\_mult\_55                   & \multicolumn{1}{l|}{0.008} & \multicolumn{1}{l|}{316.941}  & \multicolumn{1}{l|}{3220.209} & \multicolumn{1}{l|}{3120.001} & TO     \\
mod5mils\_65                    & \multicolumn{1}{l|}{0.07}  & \multicolumn{1}{l|}{1.77}     & \multicolumn{1}{l|}{73.907}   & \multicolumn{1}{l|}{84.206}   & 1298.104 \\
or                              & \multicolumn{1}{l|}{0.004} & \multicolumn{1}{l|}{0.252}    & \multicolumn{1}{l|}{3.237}    & \multicolumn{1}{l|}{5.847}    & 99.95    \\
qaoa5                           & \multicolumn{1}{l|}{0.004} & \multicolumn{1}{l|}{0.096}    & \multicolumn{1}{l|}{0.446}    & \multicolumn{1}{l|}{5.856}    & 54.485   \\
qft\_8                          & \multicolumn{1}{l|}{0.009} & \multicolumn{1}{l|}{TO}     & \multicolumn{1}{l|}{TO}     & \multicolumn{1}{l|}{TO}     & TO     \\
rc\_adder\_6                    & \multicolumn{1}{l|}{0.013} & \multicolumn{1}{l|}{TO}     & \multicolumn{1}{l|}{TO}     & \multicolumn{1}{l|}{TO}     & TO     \\
tof\_4                          & \multicolumn{1}{l|}{0.006} & \multicolumn{1}{l|}{1.912}    & \multicolumn{1}{l|}{30.767}   & \multicolumn{1}{l|}{18.029}   & 799.455  \\
tof\_5                          & \multicolumn{1}{l|}{0.006} & \multicolumn{1}{l|}{2.19}     & \multicolumn{1}{l|}{293.749}  & \multicolumn{1}{l|}{274.756}  & TO     \\
toffoli                         & \multicolumn{1}{l|}{0.004} & \multicolumn{1}{l|}{0.177}    & \multicolumn{1}{l|}{0.715}    & \multicolumn{1}{l|}{4.613}    & 35.449   \\
vbe\_adder\_3                   & \multicolumn{1}{l|}{0.008} & \multicolumn{1}{l|}{36.666}   & \multicolumn{1}{l|}{987.089}  & \multicolumn{1}{l|}{555.354}  & TO     \\
vqe\_8\_0\_5\_100               & \multicolumn{1}{l|}{0.009} & \multicolumn{1}{l|}{3547.93}  & \multicolumn{1}{l|}{2790.169} & \multicolumn{1}{l|}{2405.887} & TO     \\
vqe\_8\_0\_10\_100              & \multicolumn{1}{l|}{0.011} & \multicolumn{1}{l|}{1737.457} & \multicolumn{1}{l|}{TO}     & \multicolumn{1}{l|}{TO}     & TO     \\
vqe\_8\_1\_5\_100               & \multicolumn{1}{l|}{0.005} & \multicolumn{1}{l|}{0.918}    & \multicolumn{1}{l|}{21.474}   & \multicolumn{1}{l|}{16.271}   & 727.958  \\
vqe\_8\_1\_10\_100              & \multicolumn{1}{l|}{0.009} & \multicolumn{1}{l|}{349.899}  & \multicolumn{1}{l|}{1437.787} & \multicolumn{1}{l|}{1668.212} & TO     \\
vqe\_8\_2\_5\_100               & \multicolumn{1}{l|}{0.008} & \multicolumn{1}{l|}{96.483}   & \multicolumn{1}{l|}{349.531}  & \multicolumn{1}{l|}{775.661}  & TO     \\
vqe\_8\_2\_10\_100              & \multicolumn{1}{l|}{0.012} & \multicolumn{1}{l|}{1601.693} & \multicolumn{1}{l|}{TO}     & \multicolumn{1}{l|}{TO}     & TO     \\
vqe\_8\_3\_5\_100               & \multicolumn{1}{l|}{0.008} & \multicolumn{1}{l|}{29.986}   & \multicolumn{1}{l|}{1496.082} & \multicolumn{1}{l|}{1136.165} & TO     \\
vqe\_8\_3\_10\_100              & \multicolumn{1}{l|}{0.012} & \multicolumn{1}{l|}{TO}     & \multicolumn{1}{l|}{TO}     & \multicolumn{1}{l|}{TO}     & TO     \\
vqe\_8\_4\_5\_100               & \multicolumn{1}{l|}{0.007} & \multicolumn{1}{l|}{16.793}   & \multicolumn{1}{l|}{803.495}  & \multicolumn{1}{l|}{761.796}  & TO     \\
vqe\_8\_4\_10\_100              & \multicolumn{1}{l|}{0.011} & \multicolumn{1}{l|}{TO}     & \multicolumn{1}{l|}{TO}     & \multicolumn{1}{l|}{TO}     & TO     \\
16QBT\_05CYC\_TFL\_0            & \multicolumn{1}{l|}{0.005} & \multicolumn{1}{l|}{0.41}     & \multicolumn{1}{l|}{0.342}    & \multicolumn{1}{l|}{6.948}    & 27.547   \\
16QBT\_10CYC\_TFL\_0            & \multicolumn{1}{l|}{0.006} & \multicolumn{1}{l|}{2.902}    & \multicolumn{1}{l|}{5.386}    & \multicolumn{1}{l|}{11.739}   & 216.8    \\
16QBT\_15CYC\_TFL\_0            & \multicolumn{1}{l|}{0.011} & \multicolumn{1}{l|}{6678.632} & \multicolumn{1}{l|}{1344.541} & \multicolumn{1}{l|}{595.658}  & TO     \\
16QBT\_20CYC\_TFL\_0            & \multicolumn{1}{l|}{0.009} & \multicolumn{1}{l|}{TO}     & \multicolumn{1}{l|}{793.442}  & \multicolumn{1}{l|}{TO}     & TO     \\
16QBT\_25CYC\_TFL\_0            & \multicolumn{1}{l|}{0.013} & \multicolumn{1}{l|}{TO}     & \multicolumn{1}{l|}{1757.849} & \multicolumn{1}{l|}{363.41}   & TO     \\
16QBT\_30CYC\_TFL\_0            & \multicolumn{1}{l|}{0.012} & \multicolumn{1}{l|}{3198.71}  & \multicolumn{1}{l|}{3548.461} & \multicolumn{1}{l|}{292.283}  & TO     \\
16QBT\_35CYC\_TFL\_0            & \multicolumn{1}{l|}{0.013} & \multicolumn{1}{l|}{5029.62}  & \multicolumn{1}{l|}{5076.226} & \multicolumn{1}{l|}{2362.786} & TO     \\
16QBT\_40CYC\_TFL\_0            & \multicolumn{1}{l|}{0.017} & \multicolumn{1}{l|}{TO}     & \multicolumn{1}{l|}{TO}     & \multicolumn{1}{l|}{TO}     & TO     \\
16QBT\_45CYC\_TFL\_0            & \multicolumn{1}{l|}{0.017} & \multicolumn{1}{l|}{TO}     & \multicolumn{1}{l|}{TO}     & \multicolumn{1}{l|}{TO}     & TO     \\ \hline
\end{tabular}
\end{table}

\begin{table}[H]
\caption{Run time for all tools on the Sycamore platform.}
\label{fig:sycamore-time}
\centering
\begin{tabular}{|l|lllll|}
\hline
\multicolumn{1}{|c|}{Sycamore} & \multicolumn{5}{c|}{Run time (s)}                                                                                                     \\ \hline
Circuit                        & \multicolumn{1}{l|}{SABRE} & \multicolumn{1}{l|}{Q-Synth}  & \multicolumn{1}{l|}{QuilLS}   & \multicolumn{1}{l|}{TB-OLSQ2} & OLSQ2    \\ \hline
4gt13\_92                      & \multicolumn{1}{l|}{0.01}  & \multicolumn{1}{l|}{5.329}    & \multicolumn{1}{l|}{TO}     & \multicolumn{1}{l|}{880.686}  & TO     \\
4mod5-v1\_22                   & \multicolumn{1}{l|}{0.007} & \multicolumn{1}{l|}{0.564}    & \multicolumn{1}{l|}{56.974}   & \multicolumn{1}{l|}{9.747}    & 1610.545 \\
adder                          & \multicolumn{1}{l|}{0.006} & \multicolumn{1}{l|}{0.065}    & \multicolumn{1}{l|}{1.193}    & \multicolumn{1}{l|}{4.038}    & 62.617   \\
barenco\_tof\_4                & \multicolumn{1}{l|}{0.009} & \multicolumn{1}{l|}{1.337}    & \multicolumn{1}{l|}{316.809}  & \multicolumn{1}{l|}{228.015}  & TO     \\
barenco\_tof\_5                & \multicolumn{1}{l|}{0.011} & \multicolumn{1}{l|}{7.901}    & \multicolumn{1}{l|}{2998.904} & \multicolumn{1}{l|}{708.275}  & TO     \\
mod\_mult\_55                  & \multicolumn{1}{l|}{0.012} & \multicolumn{1}{l|}{151.213}  & \multicolumn{1}{l|}{3509.841} & \multicolumn{1}{l|}{990.915}  & TO     \\
mod5mils\_65                   & \multicolumn{1}{l|}{0.007} & \multicolumn{1}{l|}{1.84}     & \multicolumn{1}{l|}{603.655}  & \multicolumn{1}{l|}{710.517}  & TO     \\
or                             & \multicolumn{1}{l|}{0.006} & \multicolumn{1}{l|}{0.206}    & \multicolumn{1}{l|}{26.062}   & \multicolumn{1}{l|}{10.439}   & 419.816  \\
qaoa5                          & \multicolumn{1}{l|}{0.006} & \multicolumn{1}{l|}{0.066}    & \multicolumn{1}{l|}{1.454}    & \multicolumn{1}{l|}{4.149}    & 79.618   \\
qft\_8                         & \multicolumn{1}{l|}{0.01}  & \multicolumn{1}{l|}{TO}     & \multicolumn{1}{l|}{TO}     & \multicolumn{1}{l|}{TO}     & TO     \\
rc\_adder\_6                   & \multicolumn{1}{l|}{0.012} & \multicolumn{1}{l|}{TO}     & \multicolumn{1}{l|}{TO}     & \multicolumn{1}{l|}{1950.9}   & TO     \\
tof\_4                         & \multicolumn{1}{l|}{0.008} & \multicolumn{1}{l|}{0.207}    & \multicolumn{1}{l|}{198.797}  & \multicolumn{1}{l|}{9.122}    & 1983.439 \\
tof\_5                         & \multicolumn{1}{l|}{0.008} & \multicolumn{1}{l|}{0.264}    & \multicolumn{1}{l|}{3461.139} & \multicolumn{1}{l|}{11.882}   & TO     \\
toffoli                        & \multicolumn{1}{l|}{0.006} & \multicolumn{1}{l|}{0.099}    & \multicolumn{1}{l|}{2.614}    & \multicolumn{1}{l|}{4.44}     & 71.593   \\
vbe\_adder\_3                  & \multicolumn{1}{l|}{0.009} & \multicolumn{1}{l|}{83.461}   & \multicolumn{1}{l|}{TO}     & \multicolumn{1}{l|}{1502.897} & TO     \\
vqe\_8\_0\_5\_100              & \multicolumn{1}{l|}{0.011} & \multicolumn{1}{l|}{232.829}  & \multicolumn{1}{l|}{TO}     & \multicolumn{1}{l|}{1445.775} & TO     \\
vqe\_8\_0\_10\_100             & \multicolumn{1}{l|}{0.014} & \multicolumn{1}{l|}{575.987}  & \multicolumn{1}{l|}{TO}     & \multicolumn{1}{l|}{6916.954} & TO     \\
vqe\_8\_1\_5\_100              & \multicolumn{1}{l|}{0.007} & \multicolumn{1}{l|}{0.581}    & \multicolumn{1}{l|}{14.48}    & \multicolumn{1}{l|}{18.133}   & 420.043  \\
vqe\_8\_1\_10\_100             & \multicolumn{1}{l|}{0.01}  & \multicolumn{1}{l|}{133.487}  & \multicolumn{1}{l|}{TO}     & \multicolumn{1}{l|}{851.404}  & TO     \\
vqe\_8\_2\_5\_100              & \multicolumn{1}{l|}{0.009} & \multicolumn{1}{l|}{4.842}    & \multicolumn{1}{l|}{913.003}  & \multicolumn{1}{l|}{589.854}  & TO     \\
vqe\_8\_2\_10\_100             & \multicolumn{1}{l|}{0.013} & \multicolumn{1}{l|}{114.877}  & \multicolumn{1}{l|}{TO}     & \multicolumn{1}{l|}{7119.92}  & TO     \\
vqe\_8\_3\_5\_100              & \multicolumn{1}{l|}{0.01}  & \multicolumn{1}{l|}{16.628}   & \multicolumn{1}{l|}{TO}     & \multicolumn{1}{l|}{2791.953} & TO     \\
vqe\_8\_3\_10\_100             & \multicolumn{1}{l|}{0.013} & \multicolumn{1}{l|}{1530.459} & \multicolumn{1}{l|}{TO}     & \multicolumn{1}{l|}{TO}     & TO     \\
vqe\_8\_4\_5\_100              & \multicolumn{1}{l|}{0.009} & \multicolumn{1}{l|}{2.913}    & \multicolumn{1}{l|}{3885.319} & \multicolumn{1}{l|}{95.447}   & TO     \\
vqe\_8\_4\_10\_100             & \multicolumn{1}{l|}{0.012} & \multicolumn{1}{l|}{TO}     & \multicolumn{1}{l|}{TO}     & \multicolumn{1}{l|}{TO}     & TO     \\
16QBT\_05CYC\_TFL\_0           & \multicolumn{1}{l|}{0.007} & \multicolumn{1}{l|}{0.148}    & \multicolumn{1}{l|}{1.085}    & \multicolumn{1}{l|}{9.758}    & 41.498   \\
16QBT\_10CYC\_TFL\_0           & \multicolumn{1}{l|}{0.01}  & \multicolumn{1}{l|}{0.2}      & \multicolumn{1}{l|}{4.232}    & \multicolumn{1}{l|}{13.679}   & 205.798  \\
16QBT\_15CYC\_TFL\_0           & \multicolumn{1}{l|}{0.009} & \multicolumn{1}{l|}{0.21}     & \multicolumn{1}{l|}{16.65}    & \multicolumn{1}{l|}{19.655}   & 516.594  \\
16QBT\_20CYC\_TFL\_0           & \multicolumn{1}{l|}{0.011} & \multicolumn{1}{l|}{0.23}     & \multicolumn{1}{l|}{109.893}  & \multicolumn{1}{l|}{23.341}   & 1121.692 \\
16QBT\_25CYC\_TFL\_0           & \multicolumn{1}{l|}{0.011} & \multicolumn{1}{l|}{0.249}    & \multicolumn{1}{l|}{214.777}  & \multicolumn{1}{l|}{24.603}   & 1633.098 \\
16QBT\_30CYC\_TFL\_0           & \multicolumn{1}{l|}{0.012} & \multicolumn{1}{l|}{0.248}    & \multicolumn{1}{l|}{357.512}  & \multicolumn{1}{l|}{26.48}    & 3414.339 \\
16QBT\_35CYC\_TFL\_0           & \multicolumn{1}{l|}{0.013} & \multicolumn{1}{l|}{0.256}    & \multicolumn{1}{l|}{83.264}   & \multicolumn{1}{l|}{28.783}   & 2670.242 \\
16QBT\_40CYC\_TFL\_0           & \multicolumn{1}{l|}{0.016} & \multicolumn{1}{l|}{0.285}    & \multicolumn{1}{l|}{132.175}  & \multicolumn{1}{l|}{32.503}   & 4419.062 \\
16QBT\_45CYC\_TFL\_0           & \multicolumn{1}{l|}{0.016} & \multicolumn{1}{l|}{0.282}    & \multicolumn{1}{l|}{1157.75}  & \multicolumn{1}{l|}{35.411}   & 6997.928 \\
54QBT\_05CYC\_QSE\_0           & \multicolumn{1}{l|}{0.012} & \multicolumn{1}{l|}{4.344}    & \multicolumn{1}{l|}{11.682}   & \multicolumn{1}{l|}{291.38}   & 213.489  \\
54QBT\_10CYC\_QSE\_0           & \multicolumn{1}{l|}{0.024} & \multicolumn{1}{l|}{1.928}    & \multicolumn{1}{l|}{21.179}   & \multicolumn{1}{l|}{TO}     & 1122.544 \\
54QBT\_15CYC\_QSE\_0           & \multicolumn{1}{l|}{0.035} & \multicolumn{1}{l|}{1.361}    & \multicolumn{1}{l|}{61.351}   & \multicolumn{1}{l|}{1068.261} & 2326.7   \\
54QBT\_20CYC\_QSE\_0           & \multicolumn{1}{l|}{0.045} & \multicolumn{1}{l|}{1.678}    & \multicolumn{1}{l|}{102.67}   & \multicolumn{1}{l|}{380.719}  & 3657.075 \\
54QBT\_25CYC\_QSE\_0           & \multicolumn{1}{l|}{0.055} & \multicolumn{1}{l|}{1.914}    & \multicolumn{1}{l|}{103.908}  & \multicolumn{1}{l|}{267.151}  & 5487.964 \\
54QBT\_30CYC\_QSE\_0           & \multicolumn{1}{l|}{0.066} & \multicolumn{1}{l|}{2.349}    & \multicolumn{1}{l|}{152.103}  & \multicolumn{1}{l|}{308.543}  & TO     \\
54QBT\_35CYC\_QSE\_0           & \multicolumn{1}{l|}{0.074} & \multicolumn{1}{l|}{2.452}    & \multicolumn{1}{l|}{225.649}  & \multicolumn{1}{l|}{203.266}  & TO     \\
54QBT\_40CYC\_QSE\_0           & \multicolumn{1}{l|}{0.083} & \multicolumn{1}{l|}{2.699}    & \multicolumn{1}{l|}{307.721}  & \multicolumn{1}{l|}{254.643}  & TO     \\
54QBT\_45CYC\_QSE\_0           & \multicolumn{1}{l|}{0.094} & \multicolumn{1}{l|}{2.891}    & \multicolumn{1}{l|}{537.394}  & \multicolumn{1}{l|}{210.628}  & TO     \\ \hline
\end{tabular}
\end{table}

\begin{table}[H]
\caption{Run time for all tools on the Aspen-M platform.}
\label{fig:aspen-time}
\centering
\begin{tabular}{|l|lllll|}
\hline
\multicolumn{1}{|c|}{Aspen-M} & \multicolumn{5}{c|}{Run time (s)}                                                                                                   \\ \hline
Circuit                       & \multicolumn{1}{l|}{SABRE} & \multicolumn{1}{l|}{Q-Synth} & \multicolumn{1}{l|}{QuilLS}   & \multicolumn{1}{l|}{TB-OLSQ2} & OLSQ2   \\ \hline
4gt13\_92                     & \multicolumn{1}{l|}{0.01}  & \multicolumn{1}{l|}{7.80}    & \multicolumn{1}{l|}{TO}     & \multicolumn{1}{l|}{373.71}   & TO \\
4mod5-v1\_22                  & \multicolumn{1}{l|}{0.01}  & \multicolumn{1}{l|}{1.96}    & \multicolumn{1}{l|}{137.293}  & \multicolumn{1}{l|}{9.09}     & 1508.54 \\
adder                         & \multicolumn{1}{l|}{0.01}  & \multicolumn{1}{l|}{0.10}    & \multicolumn{1}{l|}{2.458}    & \multicolumn{1}{l|}{4.99}     & 75.48   \\
barenco\_tof\_4               & \multicolumn{1}{l|}{0.01}  & \multicolumn{1}{l|}{10.31}   & \multicolumn{1}{l|}{5272.083} & \multicolumn{1}{l|}{335.20}   & TO \\
barenco\_tof\_5               & \multicolumn{1}{l|}{0.01}  & \multicolumn{1}{l|}{44.57}   & \multicolumn{1}{l|}{TO}     & \multicolumn{1}{l|}{1413.26}  & TO \\
mod\_mult\_55                 & \multicolumn{1}{l|}{0.02}  & \multicolumn{1}{l|}{112.78}  & \multicolumn{1}{l|}{TO}     & \multicolumn{1}{l|}{5717.58}  & TO \\
mod5mils\_65                  & \multicolumn{1}{l|}{0.01}  & \multicolumn{1}{l|}{2.68}    & \multicolumn{1}{l|}{1164.662} & \multicolumn{1}{l|}{73.60}    & TO \\
or                            & \multicolumn{1}{l|}{0.01}  & \multicolumn{1}{l|}{0.47}    & \multicolumn{1}{l|}{39.829}   & \multicolumn{1}{l|}{7.50}     & 465.47  \\
qaoa5                         & \multicolumn{1}{l|}{0.01}  & \multicolumn{1}{l|}{0.11}    & \multicolumn{1}{l|}{2.839}    & \multicolumn{1}{l|}{4.61}     & 96.00   \\
qft\_8                        & \multicolumn{1}{l|}{0.01}  & \multicolumn{1}{l|}{TO} & \multicolumn{1}{l|}{TO}     & \multicolumn{1}{l|}{TO}  & TO \\
rc\_adder\_6                  & \multicolumn{1}{l|}{0.02}  & \multicolumn{1}{l|}{TO} & \multicolumn{1}{l|}{TO}     & \multicolumn{1}{l|}{218.64}   & TO \\
tof\_4                        & \multicolumn{1}{l|}{0.01}  & \multicolumn{1}{l|}{0.35}    & \multicolumn{1}{l|}{3458.841} & \multicolumn{1}{l|}{10.57}    & TO \\
tof\_5                        & \multicolumn{1}{l|}{0.01}  & \multicolumn{1}{l|}{0.49}    & \multicolumn{1}{l|}{TO}     & \multicolumn{1}{l|}{13.70}    & TO \\
toffoli                       & \multicolumn{1}{l|}{0.01}  & \multicolumn{1}{l|}{0.19}    & \multicolumn{1}{l|}{5.375}    & \multicolumn{1}{l|}{3.67}     & 67.46   \\
vbe\_adder\_3                 & \multicolumn{1}{l|}{0.01}  & \multicolumn{1}{l|}{164.17}  & \multicolumn{1}{l|}{TO}     & \multicolumn{1}{l|}{766.23}   & TO \\
vqe\_8\_0\_5\_100             & \multicolumn{1}{l|}{0.01}  & \multicolumn{1}{l|}{2641.92} & \multicolumn{1}{l|}{TO}     & \multicolumn{1}{l|}{TO}  & TO \\
vqe\_8\_0\_10\_100            & \multicolumn{1}{l|}{0.02}  & \multicolumn{1}{l|}{3574.55} & \multicolumn{1}{l|}{TO}     & \multicolumn{1}{l|}{TO}  & TO \\
vqe\_8\_1\_5\_100             & \multicolumn{1}{l|}{0.01}  & \multicolumn{1}{l|}{0.97}    & \multicolumn{1}{l|}{150.103}  & \multicolumn{1}{l|}{23.86}    & 3243.88 \\
vqe\_8\_1\_10\_100            & \multicolumn{1}{l|}{0.01}  & \multicolumn{1}{l|}{1144.67} & \multicolumn{1}{l|}{TO}     & \multicolumn{1}{l|}{2573.55}  & TO \\
vqe\_8\_2\_5\_100             & \multicolumn{1}{l|}{0.01}  & \multicolumn{1}{l|}{59.25}   & \multicolumn{1}{l|}{4052.27}  & \multicolumn{1}{l|}{898.78}   & TO \\
vqe\_8\_2\_10\_100            & \multicolumn{1}{l|}{0.02}  & \multicolumn{1}{l|}{6074.12} & \multicolumn{1}{l|}{TO}     & \multicolumn{1}{l|}{TO}  & TO \\
vqe\_8\_3\_5\_100             & \multicolumn{1}{l|}{0.01}  & \multicolumn{1}{l|}{25.15}   & \multicolumn{1}{l|}{TO}     & \multicolumn{1}{l|}{1434.67}  & TO \\
vqe\_8\_3\_10\_100            & \multicolumn{1}{l|}{0.02}  & \multicolumn{1}{l|}{TO} & \multicolumn{1}{l|}{TO}     & \multicolumn{1}{l|}{TO}  & TO \\
vqe\_8\_4\_5\_100             & \multicolumn{1}{l|}{0.01}  & \multicolumn{1}{l|}{32.56}   & \multicolumn{1}{l|}{TO}     & \multicolumn{1}{l|}{189.64}   & TO \\
vqe\_8\_4\_10\_100            & \multicolumn{1}{l|}{0.02}  & \multicolumn{1}{l|}{TO} & \multicolumn{1}{l|}{TO}     & \multicolumn{1}{l|}{TO}  & TO \\
16QBT\_05CYC\_TFL\_0          & \multicolumn{1}{l|}{0.01}  & \multicolumn{1}{l|}{0.24}    & \multicolumn{1}{l|}{2.168}    & \multicolumn{1}{l|}{12.61}    & 51.29   \\
16QBT\_10CYC\_TFL\_0          & \multicolumn{1}{l|}{0.01}  & \multicolumn{1}{l|}{0.34}    & \multicolumn{1}{l|}{10.414}   & \multicolumn{1}{l|}{23.82}    & 721.11  \\
16QBT\_15CYC\_TFL\_0          & \multicolumn{1}{l|}{0.01}  & \multicolumn{1}{l|}{0.37}    & \multicolumn{1}{l|}{42.271}   & \multicolumn{1}{l|}{26.05}    & 715.58  \\
16QBT\_20CYC\_TFL\_0          & \multicolumn{1}{l|}{0.02}  & \multicolumn{1}{l|}{0.40}    & \multicolumn{1}{l|}{108.668}  & \multicolumn{1}{l|}{62.14}    & 4877.86 \\
16QBT\_25CYC\_TFL\_0          & \multicolumn{1}{l|}{0.02}  & \multicolumn{1}{l|}{0.45}    & \multicolumn{1}{l|}{153.245}  & \multicolumn{1}{l|}{33.30}    & TO \\
16QBT\_30CYC\_TFL\_0          & \multicolumn{1}{l|}{0.02}  & \multicolumn{1}{l|}{0.40}    & \multicolumn{1}{l|}{407.181}  & \multicolumn{1}{l|}{36.78}    & TO \\
16QBT\_35CYC\_TFL\_0          & \multicolumn{1}{l|}{0.02}  & \multicolumn{1}{l|}{0.45}    & \multicolumn{1}{l|}{412.318}  & \multicolumn{1}{l|}{41.89}    & TO \\
16QBT\_40CYC\_TFL\_0          & \multicolumn{1}{l|}{0.02}  & \multicolumn{1}{l|}{0.49}    & \multicolumn{1}{l|}{270.241}  & \multicolumn{1}{l|}{55.95}    & TO \\
16QBT\_45CYC\_TFL\_0          & \multicolumn{1}{l|}{0.02}  & \multicolumn{1}{l|}{0.48}    & \multicolumn{1}{l|}{1565.592} & \multicolumn{1}{l|}{48.08}    & TO \\
54QBT\_05CYC\_QSE\_0          & \multicolumn{1}{l|}{0.02}  & \multicolumn{1}{l|}{24.76}   & \multicolumn{1}{l|}{138.734}  & \multicolumn{1}{l|}{229.38}   & TO \\
54QBT\_10CYC\_QSE\_0          & \multicolumn{1}{l|}{0.03}  & \multicolumn{1}{l|}{TO} & \multicolumn{1}{l|}{TO}     & \multicolumn{1}{l|}{TO}  & TO \\
54QBT\_15CYC\_QSE\_0          & \multicolumn{1}{l|}{0.04}  & \multicolumn{1}{l|}{TO} & \multicolumn{1}{l|}{TO}     & \multicolumn{1}{l|}{TO}  & TO \\
54QBT\_20CYC\_QSE\_0          & \multicolumn{1}{l|}{0.05}  & \multicolumn{1}{l|}{TO} & \multicolumn{1}{l|}{TO}     & \multicolumn{1}{l|}{TO}  & TO \\
54QBT\_25CYC\_QSE\_0          & \multicolumn{1}{l|}{0.07}  & \multicolumn{1}{l|}{TO} & \multicolumn{1}{l|}{TO}     & \multicolumn{1}{l|}{TO}  & TO \\
54QBT\_30CYC\_QSE\_0          & \multicolumn{1}{l|}{0.07}  & \multicolumn{1}{l|}{TO} & \multicolumn{1}{l|}{TO}     & \multicolumn{1}{l|}{TO}  & TO \\
54QBT\_35CYC\_QSE\_0          & \multicolumn{1}{l|}{0.08}  & \multicolumn{1}{l|}{TO} & \multicolumn{1}{l|}{TO}     & \multicolumn{1}{l|}{TO}  & TO \\
54QBT\_40CYC\_QSE\_0          & \multicolumn{1}{l|}{0.09}  & \multicolumn{1}{l|}{TO} & \multicolumn{1}{l|}{TO}     & \multicolumn{1}{l|}{TO}  & TO \\
54QBT\_45CYC\_QSE\_0          & \multicolumn{1}{l|}{0.11}  & \multicolumn{1}{l|}{TO} & \multicolumn{1}{l|}{TO}     & \multicolumn{1}{l|}{TO}  & TO \\ \hline
\end{tabular}
\end{table}

\end{document}